\documentclass[usenatbib,useAMS,letterpaper]{mnras}
\usepackage{float}
\usepackage{graphicx}
\usepackage{epstopdf}
\usepackage{ textcomp}
\usepackage{multirow}
\usepackage{ctable}
\usepackage{url}
\usepackage{times}
\usepackage{amsmath}
\usepackage{amssymb}
\usepackage{xspace}
\usepackage{mathrsfs} 
\usepackage{tabularx}
\usepackage{fixltx2e}
\usepackage{color}
\usepackage{placeins}
\usepackage{comment}
\usepackage{enumerate}

\newcommand{\acknowledgments}{\begin{small}\section*{Acknowledgements}\end{small}}

\defcitealias{2019MNRAS.487.4393S}{Paper {\small I}}
\defcitealias{2020MNRAS.491.1190S}{Paper {\small II}}
\newcommand\sref[1]{\hyperref[#1]{\S~\ref*{#1}}}
\newcommand\fref[1]{\hyperref[#1]{Fig.~\ref*{#1}}}
\newcommand\Eqref[1]{Eq.~(\hyperref[#1]{\ref*{#1}})}
\newcommand\eeqref[1]{Eqs.~(\hyperref[#1]{\ref*{#1}})}

\newcommand\tref[1]{\hyperref[#1]{Table~\ref*{#1}}}
\newcommand\aref[1]{\hyperref[#1]{Appendix~\ref*{#1}}}

\newcommand{\ksu}[1]{\textcolor{black}{#1}}

\newcommand{\hlt}[1]{\textcolor{blue}{#1}}

\newcommand{\oneline}[1]{%
  \newdimen{\namewidth}%
  \setlength{\namewidth}{\widthof{#1}}%
  \ifthenelse{\lengthtest{\namewidth < \textwidth}}%
  {#1}
  {\resizebox{\textwidth}{!}{#1}}
}
 
\title[]{Self-regulation of high-redshift black hole accretion via jets: challenges for SMBH formation}
\author[]{
\parbox[t]{\textwidth}{
Kung-Yi Su$^{1}$\thanks{E-mail: kungyisu@g.harvard.edu},  Greg L. Bryan$^{2}$, Zolt\'{a}n Haiman$^{2}$},  
\vspace*{6pt} \\
$^1$Black Hole Initiative, Harvard University, 20 Garden Street, Cambridge, MA 02138, USA\\
$^2$Department of Astronomy, Columbia University, 550 West 120th Street, New York, NY 10027, USA\\
}
\usepackage[all]{hypcap}

\begin{document}
\long\def\/*#1*/{}
\date{Submitted to MNRAS}

\pagerange{\pageref{firstpage}--\pageref{lastpage}} \pubyear{2021}

\maketitle

\label{firstpage}

\begin{abstract}
The early growth of black holes (BHs) in atomic-cooling halos is likely influenced by feedback on the surrounding gas. While the effects of radiative feedback are well-documented, mechanical feedback, particularly from AGN jets, has been comparatively less explored. Building on our previous work that examined the growth of a 100 ${\rm M_\odot}$ BH in a constant density environment regulated by AGN jets, we expand the initial BH mass range from 1 to $10^4$ ${\rm M_\odot}$ and adopt a more realistic density profile for atomic-cooling halos. We reaffirm the validity of our analytic models for jet cocoon propagation and feedback regulation. We identify several critical radii—namely, the terminal radius of jet cocoon propagation, the isotropization radius of the jet cocoon, and the core radius of the atomic-cooling halo—that are crucial in determining BH growth given specific gas properties and jet feedback parameters. In a significant portion of the parameter space, our findings show that jet feedback substantially disrupts the halo’s core during the initial feedback episode, preventing BH growth beyond $10^4 \, {\rm M_\odot}$. Conversely, conditions characterized by low jet velocities and high gas densities enable sustained BH growth over extended periods. We provide a prediction for the black hole mass growth as a function of time and feedback parameters. We found that, to form a supermassive BH ($>10^6 {\rm M_\odot}$) within 1 Gyr entirely by accreting gas from an atomic-cooling halo, the jet energy feedback efficiency must be $\lesssim 10^{-4} \dot{M}_{\rm BH} c^2$ even if the seed BH mass is $10^4 {\rm M_\odot}$.
\end{abstract}


\section{Introduction}
\label{S:intro}

Active galactic nucleus (AGN) feedback is a crucial factor in the evolution of galaxies, particularly in the suppression of star formation in massive galaxies and clusters, thus maintaining their ``red and dead'' status for a substantial part of cosmic history. Research extensively indicates that AGN jet feedback mechanisms are theoretically able to quench star formation and arrest cooling flows within galaxy-scale simulations \citep[e.g.,][]{2010MNRAS.409..985D,2012MNRAS.424..190G,2012MNRAS.427.1614Y,2014ApJ...789...54L,2015ApJ...811...73L,2015ApJ...811..108P,2016ApJ...818..181Y,2017ApJ...834..208R,2017MNRAS.472.4707B,2020MNRAS.491.1190S}. Observational data also supports the notion that AGNs can provide energy outputs on par with cooling rates \citep{2004ApJ...607..800B}. Moreover, AGNs are observed ejecting gas from galaxies, contributing thermal energy through shocks, sound waves, photo-ionization, Compton heating, or by enhancing turbulence in the circumgalactic medium (CGM) and intra-cluster medium (ICM), leading to the formation of hot plasma ``bubbles'' with significant relativistic components around massive galaxies \citep[see, e.g.,][for a detailed review]{2012ARA&A..50..455F,2018ARA&A..56..625H}. In \cite{2021MNRAS.507..175S,2023arXiv231017692S}, a comprehensive parameter study on AGN jets within $10^{12}-10^{15} {\rm M_\odot}$ clusters identified certain models that can produce a sufficiently large cocoon with an extended cooling period, enabling these jets to effectively quench the galaxy.

Beyond the established instances of supermassive black holes (SMBHs) in large galaxies, the influence of AGN feedback extends to smaller dwarf galaxies and stellar-mass to intermediate-mass black holes (IMBHs) with masses ranging from $M_{\rm BH}\sim {\rm a\, few} -10^5 {\rm M_\odot}$ \citep[e.g.,][]{2017ApJ...845...50N,2018ApJ...861...50B,2018MNRAS.476..979P,2019ApJ...884..180D,2019ApJ...884...54M}. These smaller black holes, some of which are the sources of AGN jets, have been observationally documented \citep[e.g.,][]{2006ApJ...636...56G,2006ApJ...646L..95W,2008ApJ...686..838W,2011AN....332..379M,2012ApJ...753..103N,2012ApJ...750L..24R,2012Sci...337..554W,2013MNRAS.436.1546M,2013MNRAS.436.3128M,2014ApJ...787L..30R,2015MNRAS.448.1893M,2018MNRAS.478.2576M,2018MNRAS.480L..74M,2019MNRAS.488..685M}. It should not be surprising that AGN feedback impacts the development of these smaller black holes, modifying the characteristics of the surrounding gas and significantly influencing their host galaxies, particularly in dwarf and early universe galaxies.
 
Studies have identified supermassive black holes (SMBH, \ksu{$M_{\rm BH}>10^{6}{\rm M_\odot}$}) in the early universe ($z\gtrsim 4$), with indications of AGN jets \citep[e.g.,][]{2021A&A...655A..95S,2022arXiv220309527S}. \ksu{Moreover, recent observations with JWST have identified a significant population of SMBHs at relatively high redshifts, beyond $z \sim 6$ and even beyond $z \sim 10$ \citep[e.g.,][]{2024Natur.627...59M,2023ApJ...953L..29L,2023Natur.619..716C,2023ApJ...959...39H,2023ApJ...954L...4K,2023ApJ...942L..17O,2023A&A...677A.145U,2024MNRAS.531.4584S}.  The feasibility of a stellar-mass $\lesssim 100{\rm M_\odot}$ black hole or even a direct-collapse black hole $\gtrsim 10^4{\rm M_\odot}$ evolving into a supermassive black hole in such a brief time-frame remains uncertain \citep[e.g.,][]{2023ApJ...957L...3P}. }Specifically, the influence of jets on the growth rate of initial black hole \ksu{seeds} poses additional questions regarding their accretion and expansion \citep[e.g.,][]{2011ApJ...739....2P,2016MNRAS.460.4122R}. The necessity for alternative mechanisms, such as runaway mergers \citep[e.g.,][]{2002ApJ...576..899P,2004ApJ...604..632G,2021MNRAS.505.2753S}, primordial black holes, or other strategies to account for the presence of supermassive black holes in the early universe remains a significant area of inquiry \citep[see][for a comprehensive review]{2020ARA&A..58...27I}.

Previous research has tackled comparable issues, using various feedback mechanisms, including radiation \citep[e.g.,][]{2009ApJ...698..766M,2011ApJ...739....2P} and stellar winds \citep[e.g.,][]{2020MNRAS.497..302T}, or by studying the development of slightly larger black holes ($>10^4 {\rm M_\odot}$) through simulations at the scale of entire galaxies with jet feedback \citep[e.g.,][]{2019MNRAS.486.3892R,2022arXiv220108766M}. 
In the study presented in \cite{2023MNRAS.520.4258S}, we explored the influence of AGN jets on the accretion processes of black holes with masses of $100 {\rm M_\odot}$ located in dense, low-metallicity gaseous environments typical of the cores of atomic-cooling halos. Additionally, we examined the dynamics of how jet-induced cocoons expand across extensive radii. Our methodology involved a systematic variation of parameters such as gas density, temperature, and the AGN feedback mechanism to ascertain the dependency of black hole accretion rates and jet propagation characteristics on these factors.

Our approach in that work involved modeling the gas environment surrounding black holes at resolutions exceeding the Bondi radius, enabling accurate estimation of the gravitational capture of gas particles. \ksu{Starting from \citet{2023MNRAS.520.4258S}}, we employed a variety of jet models to investigate their influence on the growth of black holes and the outward propagation of jets. Through that research, we developed an analytical model that aligned with our simulation outcomes, offering insights into jet propagation and the self-regulation mechanisms of AGN fluxes and black hole accretion rates. Moreover, we predicted conditions under which super-Eddington accretion occurs. Yet, that model has limitations, being only validated in simulations with initially static gas environments and at a $100 {\rm M_\odot}$ BH mass. Indeed, as a black hole accrues mass, the assumption of constant density up to the Bondi radius becomes increasingly unrealistic within the typical size of an atomic-cooling halo. Building on this previous effort \citep{2023MNRAS.520.4258S}, this study broadens our exploration to include a wider array of black hole \ksu{seed masses from stellar-mass to the direct-collapse black hole range ($1-10^4 M_\odot$). We also included a more realistic density profile resembling the core of an atomic-cooling halo.}

In \sref{S:methods}, we summarize our initial conditions (ICs), black hole accretion model, and the AGN jet parameters we survey, as well as describe our numerical simulations. In \sref{S:toy}, we outline the toy model that describes jet propagation and self-regulation in a constant-density environment. We present the simulation results in \sref{s:result}. We then generalize the toy model to include cases where the jet cocoon propagates into a decaying density profile and further suppresses the gas density in \sref{sec:model_for_supression}.  Based on the toy model, we predict the black hole mass as a function of time, seed black hole mass, feedback mass loading, and jet velocity in \sref{sec:predict}. We discuss the limitations of the work and its observational implications in \sref{sec:discussion}. Finally, we summarize our main conclusion in \sref{sec:summary}.

\section{Methodology} \label{S:methods}

Our study involves conducting simulations on a gas box influenced by jet feedback from a $100 {\rm M_\odot}$ black hole, utilizing the {\sc GIZMO} code\footnote{A public version of this code is available at \href{http://www.tapir.caltech.edu/~phopkins/Site/GIZMO.html}{\textit{http://www.tapir.caltech.edu/$\sim$phopkins/Site/GIZMO.html}}} \citep{2015MNRAS.450...53H} in its meshless finite mass (MFM) configuration. This method combines the strengths of grid-based and smoothed-particle hydrodynamics (SPH) techniques in a Lagrangian mesh-free Godunov approach. We detail the numerical implementation and conduct thorough tests, as described in methodological papers focusing on hydrodynamics and self-gravity \citep{2015MNRAS.450...53H}. Our simulations incorporate the FIRE-2 model for cooling processes (ranging from $10-10^{10}$K), encompassing photo-electric and photo-ionization heating, along with collisional, Compton, fine structure, recombination, atomic, and molecular cooling effects. We set a lower limit for the temperature, $T_\infty$, defined in the initial conditions, to prevent the gas from cooling below this threshold due to other feedback mechanisms not covered in our simulations. In the absence of such a limit, and considering the densities in our study, molecular cooling could reduce the temperature of all gas to 10K in less than 1000 years\ksu{, even at extremely low metallicity levels ($10^{-4}$ ${\rm Z_{\odot}}$ in this case).}

\subsection{Initial conditions}

In an optimal setting, we would model black hole accretion within a cosmological simulation that precisely resolves gas dynamics at early cosmic times, similar to studies conducted on minihalos \citep{2009ApJ...701L.133A}. However, due to significant uncertainties in the conditions at high redshifts, and in order to gain a better understanding of the physics,  we adopt a simplified approach that simulates the environment near the black hole as a uniform gas patch, akin to the core of an atomic-cooling halo. We systematically vary the properties of this gas patch to explore their influence on black hole regulation. Our initial setup is a 3D box populated with randomly placed gas particles at a fixed temperature, denoted as $T_\infty$. The gas density initially adheres to a specific profile as outlined in \tref{tab:run}. For simulations with black holes less than $100 {\rm M_\odot}$, the density ($n_{\rm Bondi}$) is constant throughout. In cases involving a $10^4 {\rm M_\odot}$ black hole, the density is uniform up to the Bondi radius ($n_{\rm Bondi}$), beyond which it follows a power-law distribution, $n\propto r^{\alpha}$, as per \citet{2019MNRAS.486.3892R}. A black hole occupies the center of the setup, with the gas's initial metallicity being set at a very low level ($10^{-4}$ $Z_{\odot}$). All elements in the initial setup are dynamic, evolving in response to gravitational, hydrodynamic, and additional baryonic physics.

To attain enhanced resolution near the black hole, crucial for capturing the dynamics of accretion and feedback processes, we implement a hierarchical super-Lagrangian refinement technique \citep{2020MNRAS.491.1190S,2021MNRAS.507..175S}. This strategy enables us to achieve a superior mass resolution in the area surrounding the z-axis, where the jet originates, as outlined in \tref{tab:run}, significantly exceeding the resolution achieved in many earlier global investigations. The resolution progressively decreases with distance from the z-axis ($r_{\rm 2d}$), in a manner approximately proportional to $r_{\rm 2d}$. These computational specifics are comprehensively documented in \tref{tab:run}. Insights from our resolution analysis are detailed in the Appendix of \cite{2023MNRAS.520.4258S}.

\subsection{BH accretion}
As highlighted in the introduction, our approach to modeling black hole accretion diverges from the conventional Bondi assumption, favoring a method that simulates the gravitational capture of gas directly onto the black hole, feeding a subgrid $\alpha$-disk \citep{2020MNRAS.497.5292T}. A gas particle is designated for accretion if it is gravitationally bound to the black hole and its estimated pericentric radius is less than the sink radius, $r_{\rm acc}$. This sink radius is defined as $3\times10^{-5} - 1.5\times10^{-4}\left(m^{\rm max}_g/1.4\times10^{-6}\right)^{-1/3}$ pc, tailored to the gas density in the vicinity of the black hole. $m^{\rm max}_g$ represents the maximum gas mass resolution for each run, as labeled in  \tref{tab:run}. More specifically, the sink radius $r_{\rm acc}$ is determined as the radius around the black hole that encloses 96 ``weighted'' neighborhood gas particles, but with the constraint that it must lie within $3\times10^{-5}-1.5\times10^{-4}\left(m^{\rm max}_g/1.4\times10^{-6}\right)^{-1/3}$ pc.

While our simulations track gas movements in close proximity to the black hole, we do not explicitly simulate the accretion disk's detailed physics. Instead, we employ a subgrid $\alpha$-disk model. Within this model, gas that is accreted contributes to the mass of the $\alpha$-disk ($M_{\rm \alpha}$), which starts at zero. The mass accumulated in the $\alpha$-disk is then transferred to the black hole at a rate determined by the equation
\begin{align}
\dot{M}_{\rm acc}= M{\rm \alpha}/t_{\rm disk},
\end{align}
where $t_{\rm disk}$ set to a constant value. This duration is based on the viscous timescale of a disk as described by \cite{1973A&A....24..337S}, assuming the disk temperature is $10^4 K$. The formula for $t_{\rm disk}$ is derived as $t_{\rm disk}\sim t_{\rm ff} \mathcal{M}^2/\alpha\sim 1000 {\rm yr} \left(M_{\rm BH}/100 {\rm M_\odot}\right) \left(\alpha/0.1\right)^{-1} \left(r_{\rm acc}/10^{-4} \rm pc\right)^{1/2}$, where $t_{\rm ff}$ denotes the free-fall time at $r_{\rm acc}$, and $\mathcal{M}$ represents the Mach number. In the appendix of \cite{2023MNRAS.520.4258S}, we delve into the effects of altering $t_{\rm disk}$ on the simulation outcomes.

\subsection{Jet models}
We implement a jet model as outlined in \cite{2021MNRAS.507..175S,2023MNRAS.520.4258S,2023MNRAS.523.1104W,2024MNRAS.532.2724S}, which involves a particle spawning technique for jet launch. This technique generates new gas cells (resolution elements) to simulate jet material, endowing these cells with predetermined mass, temperature, and velocity to define the jet's specific energy. This approach grants us enhanced control over jet characteristics, as it minimizes dependence on surrounding gas conditions through neighbor-finding algorithms\footnote{{Conventionally, energy and momentum are distributed to gas particles identified via a neighbor search from the black hole, making the impact reliant on local gas attributes and their spatial arrangement. Refer to \cite{2022arXiv220306201W} for a discussion on various methodologies.}}. It also allows for the enforcement of higher resolution within jet structures, facilitating the precise simulation of light jets.

The resolution of the spawned gas particles is given in \tref{tab:run}. These particles are restricted from merging back into larger gas elements until their velocity drops below 10\% of the initial launch speed. To ensure exact conservation of linear momentum, two particles are simultaneously emitted in opposite directions along the z-axis each time the accumulated mass flux of the jet doubles the mass of a target spawned particle. The initial positioning of these particles is random, within a sphere of radius $r_0$, which is the lesser of $10^{-5}\left(m^{\rm max}_g/1.4\times10^{-6}\right)^{-1/3}$ pc or half the distance from the black hole to the nearest gas particle.

For a particle initialized at coordinates $(r_0,\theta_0,\phi_0)$ within a jet model with an opening angle $\theta_{\rm op}=1^o$, the initial velocity direction is adjusted to $2\theta_{\rm op}\theta_0/\pi$ for $\theta_0<\pi/2$ and to $\pi-2\theta_{\rm op}(\pi-\theta_0)/\pi$ for $\theta_0>\pi/2$. This configuration ensures that the trajectories of any two particles will not cross, maintaining a coherent jet structure.

We define the jet mass flux using a fixed feedback mass loading as follows:
\begin{align}
\dot{M}_{\rm jet}=\eta_{\rm m, fb}\dot{M}_{\rm BH},
\end{align}
which allows us to calculate the feedback energy and momentum fluxes of the jet through the equations:
\begin{align}
\dot{E}_{\rm jet}&=\eta_{\rm m, fb}\dot{M}_{\rm BH}  \left(\frac{1}{2} V_{\rm jet}^2+ \frac{3kT}{2\mu}  \right),\notag\\
\dot{P}_{\rm jet}&=\eta_{\rm m, fb}\dot{M}_{\rm BH}  V_{\rm jet} ,
\end{align}
where $V_{\rm jet}$ represents the chosen velocity for the jet, and $\mu$ signifies the average mass of a jet particle.

The efficiency jet energy efficiency is 
\begin{align}\label{eq:e_eff}
\eta_{\rm eff}\equiv \frac{\dot{E}_{\rm jet}}{ \dot{M}_{\rm BH} c^2}=\frac{\eta_{\rm m, fb} V_{\rm jet}^2}{2 c^2} + \frac{\eta_{\rm m, fb} 3kT}{2 \mu c^2}.
\end{align}

\begin{table*}
\begin{center}
 \caption{Physics variations of all simulations}
 \label{tab:run}
\resizebox{17.7cm}{!}
{
\begin{tabular}{c|c|cccc|ccc|ccc|cccc}
\hline
\hline
&&\multicolumn{4}{c|}{Numerical details} & \multicolumn{3}{c|}{Feedback parameters} & \multicolumn{3}{c|}{Background gas} & \multicolumn{4}{c}{Resulting averaged accretion rate and fluxes}\\
\hline
Model & $M_{\rm BH}$         & $\Delta T$ & Box size & $m^{\rm max}_{\rm g}$ & $m_{\rm jet}$  & $\eta_{\rm m, fb}$ & $V_{\rm jet}$ & $T_{\rm jet}$ & $n_{\rm Bondi}$ & $n_{index}$ & $T_\infty$ & $\langle \dot{M}_{\rm BH} \rangle$ & $\frac{\langle \dot{M}_{\rm BH} \rangle}{\dot{M}_{\rm Bondi}}$ &  $\frac{\langle \dot{M}_{\rm BH} \rangle}{\dot{M}_{\rm Edd}}$  &  $\frac{\langle \dot{E}_{\rm jet} \rangle}{\dot{E}_{\rm Edd}}$ \\
           &  ${\rm M_\odot}$   & kyr        & pc       & ${\rm M_\odot}$             & ${\rm M_\odot}$      &                    & km s$^{-1}$   & K             & cm$^{-3}$&  & K & ${\rm M_\odot} {\rm yr}^{-1}$ \\ 
\hline
\multicolumn{13}{c}{}\\[-0.2cm]
\multicolumn{13}{c}{\bf \hlt{ \large$1 {\rm M_\odot}$}}\\
\hline
\multicolumn{1}{c|}{\bf \hlt{Fiducial}}& &\multicolumn{4}{c|}{}& \multicolumn{3}{c|}{}&\multicolumn{3}{c|}{}\\
m1-$\eta$5e-1--$v$j1e4--n1e5--T1e4&1 & 1 & 0.004 & 1.4e-12 & 1e-13 &0.5 &1e4 &1e4 & 1e5 & 0 &6e3 &4.7e-13 & 1.4e-4 &2.1e-5 &5.9e-8
\\
\hline 
\multicolumn{1}{c|}{\bf \hlt{Jet velocity \& FB mass fraction}}& &\multicolumn{4}{c|}{}& \multicolumn{3}{c|}{}&\multicolumn{3}{c|}{}\\
m1-$\eta$5e-1--$v$j3e3--n1e5--T1e4 & 1&1 & 0.004 & 1.4e-12 & 1e-13 &0.5 &3e3 &1e4 & 1e5&0 &6e3&7.9e-12&2.4e-3 &3.5e-4 & 8.9e-8 \\ 
m1-$\eta$1e2--$v$j3e3--n1e5--T1e4 & 1&1 & 0.004 & 1.4e-12 & 1e-13 &100 &3e3 &1e4 & 1e5&0 &6e3&2.5e-14 & 7.7e-6 &1.1e-6&5.6e-8\\ 
m1-$\eta$5e-2--$v$j1e4--n1e5--T1e4 & 1&1 & 0.004 & 1.4e-12 & 1e-13 &0.05 &1e4 &1e4 & 1e5&0 &6e3 &1.8e-12&5.6e-4&8.3e-5&2.3e-8\\ 
m1-$\eta$1e2--$v$j1e4--n1e5--T1e4 & 1&1 & 0.004 & 1.4e-12 & 1e-13 &100 &1e4 &1e4 & 1e5&0 &6e3&2.9e-15&9.1e-7&1.3e-7&7.4e-8\\ 
m1-$\eta$5e-2--$v$j3e4--n1e5--T1e4 & 1&1 & 0.004 & 1.4e-12 & 1e-13 &0.05 &3e4 &1e4 & 1e5&0 &6e3&5.7e-13&1.7e-4&2.5e-5&6.4e-8\\ 
m1-$\eta$5e-1--$v$j3e4--n1e5--T1e4 & 1&1 & 0.004 & 1.4e-12 & 1e-13 &0.5 &3e4 &1e4 & 1e5&0 &6e3&6.1e-14&1.9e-5&2.7e-6&6.9e-8\\ 
\hline
\multicolumn{1}{c|}{\bf \hlt{Gas density}}& &\multicolumn{4}{c|}{}& \multicolumn{3}{c|}{}&\multicolumn{3}{c|}{}\\
m1-$\eta$5e-2--$v$j1e4--n1e2--T1e4 & 1& 1 & 0.004 & 1.4e-15 & 1e-16 &0.05 &1e4 &1e4 & 1e2& 0 &6e3 &8.1e-16 &2.5e-4&3.7e-8 & 1.0e-11
\\
m1-$\eta$5e-2--$v$j1e4--n1e3--T1e4 & 1& 1 & 0.004 & 1.4e-14 & 1e-15 &0.05 &1e4 &1e4 & 1e3& 0 &6e3 &1.1e-14 &3.3e-4 &4.8e-7 &1.3e-10
\\
m1-$\eta$5e-1--$v$j1e4--n1e4--T1e4 & 1& 1 & 0.004 & 1.4e-13 & 1e-14 &0.5 &1e4 &1e4 & 1e4& 0 &6e3&3.7e-14 &1.1e-4 &1.7e-6 &4.6e-9
\\
m1-$\eta$5e-1--$v$j1e4--n1e6--T1e4 & 1& 1 & 0.008 & 1.4e-11 & 1e-12 &0.5 &1e4 &1e4 & 1e6& 0 &6e3 &8.0e-12 &2.4e-4 &3.6e-4&1.0e-6
\\
\hline
\multicolumn{1}{c|}{\bf \hlt{Temperature}}&& \multicolumn{4}{c|}{}& \multicolumn{3}{c|}{}&\multicolumn{3}{c|}{}\\
m1-$\eta$5e-1--$v$j1e4--n1e5--T1e3 & 1&1 & 0.032 & 1.4e-12 & 1e-13 &0.5 &1e4 &1e3 & 1e5&0 &6e2& 1.7e-13 &3.7e-6 &7.6e-6&2.1e-8
\\
m1-$\eta$5e-1--$v$j1e4--n1e5--T1e5 & 1&0.16 & 0.0004 & 1.4e-12 & 1e-16 &0.5 &1e4 &1e3 & 1e5&0 &6e4 &3.6e-14 &2.3e-3&1.6e-6&4.5e-9
\\

\hline 
\multicolumn{13}{c}{}\\[-0.2cm]
\multicolumn{13}{c}{\bf \hlt{\large $100 {\rm M_\odot}$}}\\
\hline
\multicolumn{1}{c|}{\bf \hlt{Fiducial}}&& \multicolumn{4}{c|}{}& \multicolumn{3}{c|}{}&\multicolumn{3}{c|}{}\\
m100-$\eta$5e-2--$v$j1e4--n1e5--T1e4 (I,II,III)$^*$&100 & 100 & 0.4 & 1.4e-6 & 1e-7 &0.05 &1e4 &1e4 & 1e5&0 &6e3 &0.7-1.5e-7&2.1-4.6e-3 &0.031-0.067 &0.87-1.9e-5\\
\hline 
\multicolumn{1}{c|}{\bf \hlt{Jet velocity}}&& \multicolumn{4}{c|}{}& \multicolumn{3}{c|}{}&\multicolumn{3}{c|}{}\\
m100-$\eta$5e-2--$v$j3e3--n1e5--T1e4 &100& 90 & 0.8 & 1.4e-6 & 1e-7 &0.05 &3e3 &1e4 & 1e5&0 &6e3 &1e-5  &0.31 &4.5 & 1.1e-4\\
m100-$\eta$5e-2--$v$j3e4--n1e5--T1e4 &100& 100 & 0.4 & 1.4e-6 & 1e-7 &0.05 &3e4 &1e4 & 1e5&0 &6e3 &8.9e-9 &2.7e-4 &4.0e-3 &1e-5 \\
\hline

\multicolumn{1}{c|}{\bf \hlt{Gas density}}&& \multicolumn{4}{c|}{}& \multicolumn{3}{c|}{}&\multicolumn{3}{c|}{}\\
m100-$\eta$5e-2--$v$j3e3--n1e2--T1e4 &100& 100 & 0.4 & 1.4e-9 & 1e-10 &0.05 &1e4 &1e4 & 1e2&0 &6e3 &1.6e-11 &4.9e-4 &7.2e-6 &2e-9\\
m100-$\eta$5e-2--$v$j3e3--n1e3--T1e4 &100& 100 & 0.4 & 1.4e-8 & 1e-9 &0.05 &1e4 &1e4 & 1e3&0 &6e3  &2.3e-10 &7.0e-4 &1.0e-4 &2.9e-8\\
m100-$\eta$5e-2--$v$j3e3--n1e4--T1e4 &100& 100 & 0.4 & 1.4e-7 & 1e-8 &0.05 &1e4 &1e4 & 1e4&0 &6e3 & 3.5e-9  &1.1e-3 &1.6e-3 &4.4e-7\\
m100-$\eta$5e-2--$v$j3e3--n1e6--T1e4 &100& 40 & 0.8 & 1.4e-5 & 1e-6 &0.05 &1e4 &1e4 & 1e6&0 &6e3 &1.3e-5    &4e-2 &5.8 &1.6e-3\\
\hline
\multicolumn{1}{c|}{\bf \hlt{Gas temperature}}&& \multicolumn{4}{c|}{}& \multicolumn{3}{c|}{}&\multicolumn{3}{c|}{}\\
m100-$\eta$5e-2--$v$j3e3--n1e5--T1e3 &100& 50 & 3.2 & 1.4e-6 & 1e-7 &0.05 &1e4 &1e4 & 1e5&0 &6e3 &1.9e-6 &4.2e-3 &0.85 &2.4e-4\\
m100-$\eta$5e-2--$v$j3e3--n1e5--T1e5 &100& 12 & 0.08 & 1e-8 & 8e-10 &0.05 &1e4 &1e4 & 1e5&0 &1e5 &4.4e-9 &6.0e-2 &2.0e-3 &5.5e-7\\
\hline
\multicolumn{13}{c}{}\\[-0.2cm]
\multicolumn{13}{c}{\bf \hlt{\large $10000 {\rm M_\odot}$}}\\
\hline
\multicolumn{1}{c|}{\bf \hlt{Fiducial}}&& \multicolumn{4}{c|}{}& \multicolumn{3}{c|}{}&\multicolumn{3}{c|}{}\\
m1e4-$\eta$5e-2--$v$j1e4--n1e5--T1e4 & 1e4& 1e4 & 1280 & 5.7e-1 & 0.04 &0.05 &1e4 &1e4 & 1e5&-2 &6e3&1.2e-2&3.8e-2&5.6e1&1.6e-2
\\
\hline
\multicolumn{1}{c|}{\bf \hlt{FB mass fraction}}&& \multicolumn{4}{c|}{}& \multicolumn{3}{c|}{}&\multicolumn{3}{c|}{}\\
m1e4-$\eta$5e-1--$v$j1e4--n1e5--T1e4 & 1e4& 1e4 & 1280 & 5.7e-1 & 0.04 &0.5 &1e4 &1e4 & 1e5&-2 &6e3&4.5e-3&1.4e-2&2.0e1&5.6e-2
\\
m1e4-$\eta$1e2--$v$j1e4--n1e5--T1e4 & 1e4& 1e4 & 1280 & 5.7e-1 & 0.04 &100 &1e4 &1e4 & 1e5&-2 &6e3 &8.8e-6&2.7e-5&4.0e-2&2.2e-2
\\
\hline
\multicolumn{1}{c|}{\bf \hlt{Jet velocity}}&& \multicolumn{4}{c|}{}& \multicolumn{3}{c|}{}&\multicolumn{3}{c|}{}\\
m1e4-$\eta$5e-1--$v$j3e3--n1e5--T1e4 & 1e4& 1100 & 1280 & 5.7e-1 & 0.04 &0.5 &3e3 &1e4 & 1e5&-2 &6e3&1.4e-1&4.1e-1&6.1e2&1.5e-1
\\
m1e4-$\eta$5e-1--$v$j1e5--n1e5--T1e4 & 1e4& 6000 & 1280 & 5.7e-1 & 0.04 &0.05 &1e5 &1e4 & 1e5&-2 &6e3&2.0e-4&6.0e-4&8.8e-1&2.5e-2
\\

\hline
\multicolumn{1}{c|}{\bf \hlt{Gas density}}&& \multicolumn{4}{c|}{}& \multicolumn{3}{c|}{}&\multicolumn{3}{c|}{}\\
m1e4-$\eta$5e-2--$v$j1e4--n1e2--T1e4 & 1e4& 9450 & 320 & 1.5e-4 & 1e-5 &0.05 &1e4 &1e4 & 1e2&-2 &6e3&7.6e-8&2.3e-4&3.4e-4&9.5e-8
\\
m1e4-$\eta$5e-2--$v$j1e4--n1e3--T1e4 & 1e4& 9910 & 320 & 1.5e-3 & 1e-4 &0.05 &1e4 &1e4 & 1e3&-2 &6e3&4.8e-6&1.5e-3&2.2e-2&6.0e-6
\\
m1e4-$\eta$5e-2--$v$j1e4--n1e4--T1e4 & 1e4& 1e4 & 320 & 1.5e-2 & 1e-3 &0.05 &1e4 &1e4 & 1e4&-2 &6e3&2.9e-4&8.8e-3&1.3&3.6e-4
\\


\hline 
\hline
\end{tabular}
}
\end{center}
\begin{flushleft}
This is a partial list of simulations studied here with different jet and background gas parameters. The columns list: 
(1) Model name: The name of each model starts with the black hole mass, followed by the feedback mass loading and the jet velocity in km s$^{-1}$. The final two numbers represent the background gas density in cm$^{-3}$ and the background gas temperature in K.
(2) $\Delta T$: Simulation duration. 
(3) Box size of the simulation.
(4) $m^{\rm max}_{\rm g}$: The highest mass resolution.
(5) $m^{\rm max}_{\rm jet}$: The mass resolution of the spawned jet particles.
(6) $\eta_{\rm m, fb}$: The feedback mass loading.
(7) $V_{\rm jet}$: The initial jet velocity at spawn.
(8) $T_{\rm jet}$: The initial jet temperature at spawn.
(9) $n_\infty$: The background gas density.
(10) $T_\infty$: The background  gas temperature.
{(11) $\langle \dot{M}_{\rm BH} \rangle$: The resulting time-averaged accretion rate.
(12) $\langle \dot{M}_{\rm BH} \rangle/\dot{M}_{\rm Bondi}$: The same value over the Bondi accretion rate.  
(13) $\langle \dot{M}_{\rm BH} \rangle/\dot{M}_{\rm Edd}$: The same value over the Eddington accretion rate ($\dot{M}_{\rm Edd}\equiv\dot{L}_{\rm Edd}/0.1 c^2$).     
(14) $\langle \dot{E}_{\rm jet} \rangle/\dot{L}_{\rm Edd}$: Jet energy flux over the Eddington luminosity.}\\
$^*$ We have run three variations of the same run with different random seeds ( labeled as I, II, III) to characterize numerical stochasticity. If not specified in the rest of this paper, we are referring to run I.
\end{flushleft}
\end{table*}

\section{Review of a simple model for jet propagation and cocoon formation } \label{S:toy}
In \cite{2023MNRAS.520.4258S}, we developed a toy model describing the jet propagation and feedback regulation. We briefly summarize it here for future verification over a wider black hole mass range (and modification for a non-uniform background in \sref{sec:model_for_supression}).

\subsection{Jet propagation} \label{sec:jet_propogation_constant}
When a jet is launched, the initial cocoon (defined here as the outer shock) can be approximately described by a cylindrical expansion, with width and height $R_{\rm cocoon}(t)$  and $Z_{\rm cocoon}(t)$. The cocoon evolution follows from momentum conservation in the z-direction and energy consideration in the lateral directions \citep[e.g.,][]{1989ApJ...345L..21B,2021MNRAS.507..175S}. With this, we can write down the  propagation of the jet cocoon propagating into a background medium with density $\rho_\infty$ as a function of time, as follows:
\begin{align}
\label{eq:V_r_time}
V_R(t)=\left(\frac{\gamma^2}{72\pi\beta^2}\right)^{1/6}\dot{M}_{\rm jet}^{1/6}\rho_c^{1/6}\rho_\infty^{-1/3}V_{\rm jet}^{1/2}t^{-1/3}\notag\\
R_{\rm cocoon}(t)=\left(\frac{81\gamma^2}{512\pi\beta^2}\right)^{1/6}\dot{M}_{\rm jet}^{1/6}\rho_c^{1/6}\rho_\infty^{-1/3}V_{\rm jet}^{1/2}t^{2/3}
\end{align}
and the time dependence of $V_z$ and $z_{\rm cocoon}$ as
\begin{align}
\label{eq:V_z_time}
V_z(t)=\left(\frac{8\beta}{9\pi\gamma}\right)^{1/3}\dot{M}_{\rm jet}^{1/3}\rho_c^{-1/6}\rho_\infty^{-1/6}t^{-2/3}\notag\\
z_{\rm cocoon}(t)=\left(\frac{24\beta}{\pi\gamma}\right)^{1/3}\dot{M}_{\rm jet}^{1/3}\rho_c^{-1/6}\rho_\infty^{-1/6}t^{1/3},
\end{align}
where $\gamma\equiv \dot{E}_{\rm expansion}/\dot{E}_{\rm kin}\propto \dot{E}_{\rm jet}/\dot{E}_{\rm kin}\equiv f_{\rm kin}^{-1}$, $\beta$ is an order-of-unity factor accounting for the non-cylindrical shape of the jet cocoon, and $\rho_c$ is the cocoon gas density.
Assuming the jet results in strong shocks, $\gamma$ in the above expression roughly follows:
\begin{align}
\gamma_{\rm super-sonic} &\sim \frac{\dot{E}_{\rm post-shock}}{\dot{E}_{\rm pre-shock}} \frac{\dot{E}_{\rm pre-shock}}{\dot{E}_{\rm jet}}\frac{\dot{E}_{\rm jet}}{\dot{E}_{\rm kin}}\notag\\
&\sim\frac{\rho_{\rm post} v_{\rm post}^3}{\rho_{\rm pre} v_{\rm pre}^3}\times(1-f_{\rm loss})f_{\rm kin}^{-1}\notag\\
&\sim \frac{1}{16} \times(1-f_{\rm loss})f_{\rm kin}^{-1}\,\,\lesssim\frac{1}{16}f_{\rm kin}^{-1},
\end{align}
where $f_{\rm loss}$ represents the fraction of energy lost through radiative cooling during propagation.

We assume that $\rho_c$, the cocoon gas density, is related to the background gas density and velocity and jet velocity as
\begin{align}
\label{eq:index}
\rho_c\propto\rho_\infty^\zeta T_\infty^\xi V_{\rm jet}^\delta.
\end{align}
where $\zeta$, $\xi$, and $\delta$ are exponents that we will determine later.

Following \eeqref{eq:V_r_time} and \Eqref{eq:V_z_time}, the lateral size grows more rapidly than the $z$-expansion and eventually becomes comparable to the propagation distance in the jet direction ($R_{\rm cocoon}/z_{\rm cocoon}\sim 1$).  This happens at a height of 
\begin{align}\label{eq:z_iso}
&z_{\rm iso}\equiv r_{\rm iso}=\left(\frac{\dot{M}_{\rm jet}}{2\pi\rho_c V_{\rm jet}}\right)^{1/2}\left(\frac{16\beta}{\gamma}\right)\notag\\
&\approx 1.3\times10^{-3} {\rm pc}\,\,\, (1-f_{\rm loss})^{-1}f_{\rm kin}\times \left(\frac{n_c}{10^{5}\,{\rm cm}^{-3} }\right)^{-1/2} \notag\\
&\left(\frac{\dot{M}_{\rm jet}}{5\times10^{-9}\,{\rm M}_\odot {\rm yr}^{-1} }\right)^{1/2}\left(\frac{V_{\rm J}}{10^4\,{\rm km\, s}^{-1} }\right)^{-1/2},
\end{align}
where $n_c$ is the gas number density of the cocoon (corresponding to $\rho_c$).
Beyond this radius, the cocoon becomes isotropic, with radius $R(t)$, and the propagation is governed solely by energy conservation with:
\begin{align}\label{eq:iso_r_time}
V_{\rm R}(t)=\left(\frac{9\gamma'\dot{M}_{\rm jet}V_{\rm jet}^2\rho_c^{1/2}}{200\pi\rho_\infty^{3/2}}\right)^{1/5}
 (t-t_{\rm iso})^{-2/5}\notag\\
R(t)=\left(\frac{125\gamma'\dot{M}_{\rm jet}V_{\rm jet}^2\rho_c^{1/2}}{216\pi\rho_\infty^{3/2}}\right)^{1/5}
 (t-t_{\rm iso})^{3/5}.
 \end{align}

\subsection{Feedback self-regulation}\label{sec:self_regulation}
In \cite{2023MNRAS.520.4258S}, we concluded that jet feedback self-regulates such that the time-averaged isotropic component of the outflowing momentum flux of the jet cocoon roughly balances the free-fall inflowing momentum flux at the Bondi radius (the ``Bondi flux''). The isotropic component is defined as the radial momentum flux, assuming that the momentum flux in all directions is the minimum of the lateral or the jet-direction components. Depending on where the cocoon isotropizes, there can be two possibilities: one when $r_{\rm iso}>r_{\rm Bondi}$ and one when $r_{\rm iso}<r_{\rm Bondi}$. The resulting mass flux approximately obeys the following expression:
\begin{align}\label{eq:mdotjet_reg}
\dot{M}_{\rm jet}\sim \begin{cases}
     \propto  M_{\rm BH}^2 \rho_\infty^{2-\zeta} \,\,T_\infty^{-\xi}\,\, V_{\rm jet}^{-3-\delta}  \text{\,\,\,\,(for\,\,} r_{\rm iso}>r_{\rm Bondi})\\
     \,\\
      \propto M_{\rm BH}^2 \rho_\infty^{(3-\zeta)/2}  T_\infty^{-(1+\xi)/2} V_{\rm jet}^{-2-\delta/2}  \text{\,\,\,\,\,(for\,\,} r_{\rm iso}<r_{\rm Bondi})
    \end{cases} 
\end{align}

The jet cocoon will be elongated at the Bondi radius if $z_{\rm iso}>r_{\rm Bondi}$ or, from \eeqref{eq:z_iso}, if 
\begin{align}\label{eq:criteria}
\dot{M}_{\rm jet} > \left(\frac{\pi \gamma^2 r_{\rm Bondi}^2}{128 \beta^2}\right) \rho_c V_{\rm jet}\sim  \rho_\infty^\zeta\,\, T_\infty^\xi V_{\rm jet}^{1+\delta}.
\end{align}
Otherwise, the cocoon isotropizes before reaching the Bondi radius.

\section{Simulation results: A strong dependence on black hole mass} \label{s:result}

We now test (and eventually generalize) these expressions with numerical simulations that push to significantly higher BH masses.

\subsection{Black hole accretion and jet fluxes}\label{sec:BH_fluxes_all}
\begin{figure*}
    \centering
    \includegraphics[width=16cm]{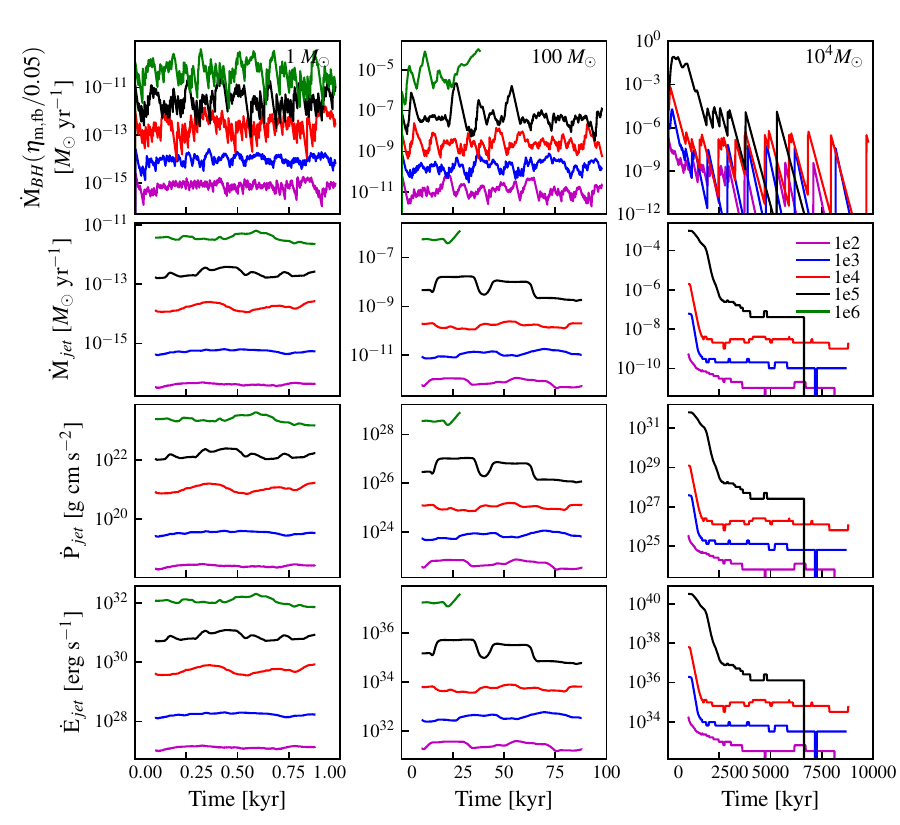}
    \caption{From top to bottom, each row of panels shows (i) the black hole accretion rate, (ii) jet mass flux, (iii) momentum flux, and (iv) energy flux, in runs, varying the initial gas density from $10^2$ to $10^6$ cm$^{-3}$ (density denoted by line color). The columns show these quantities for initial black hole mass of $1 {\rm M_\odot}$ (left), $100 {\rm M_\odot}$ (center), and $10^4 {\rm M_\odot}$ (right). The 2nd, 3rd, and 4th rows (from the top) show moving averages around the specific time of the run. Jets emanating from black holes in denser environments self-regulate to achieve higher mass, momentum, and energy flux, all of which increase super-linearly with the density. Simulations involving a $10^4 {\rm M_\odot}$ black hole initially exhibit a similar pattern; however, the accretion rates and fluxes rapidly decline thereafter. }
    \label{fig:flux_bh}
\end{figure*}
\fref{fig:flux_bh} shows the black hole accretion rate as well as the jet mass, momentum, and energy fluxes for our simulations with varying initial gas densities and black hole masses. It is clear that for all black hole masses surveyed, jets launched from black holes in denser environments self-regulate to achieve higher mass, momentum, and energy fluxes, all of which increase super-linearly with the density. However, while simulations involving a $10^4 {\rm M_\odot}$ black hole initially exhibit a similar pattern, the accretion rates and fluxes rapidly decline thereafter. The majority of the black hole accretion occurs within the first 2 Myr, after which the contribution to the net accreted mass becomes exponentially small throughout the simulations. As we will discuss in more detail below, this is primarily due to the overall drop in density profiles in some of the runs when the jet cocoon propagates beyond the core radius, approximately 1 pc. We will discuss the conditions for such density suppression and its effects in later sections.

\subsection{Agreement with analytic models for black holes up to $M_{\rm BH}<10^4\,{\rm M_\odot}$}

In this work, we survey a broader range of black hole masses, from $1-10^4 {\rm M_\odot}$. As we mentioned in \sref{sec:BH_fluxes_all}, and will return to later, a suppression of the density profile can occur for $M_{\rm BH} \gtrsim 10^4 {\rm M_\odot}$.  In this section, we show that the analytic models described in  \sref{S:toy} hold for black holes up to $M_{\rm BH} < 10^4 {\rm M_\odot}$. The case of $M_{\rm BH} = 100 {\rm M_\odot}$ was demonstrated in our previous papers, so this section focuses primarily on the $M_{\rm BH} = 1 {\rm M_\odot}$ cases.

\subsubsection{The self-regulation of the cocoon by its isotropic momentum flux}
\begin{figure*}
    \centering
    \includegraphics[width=16cm]{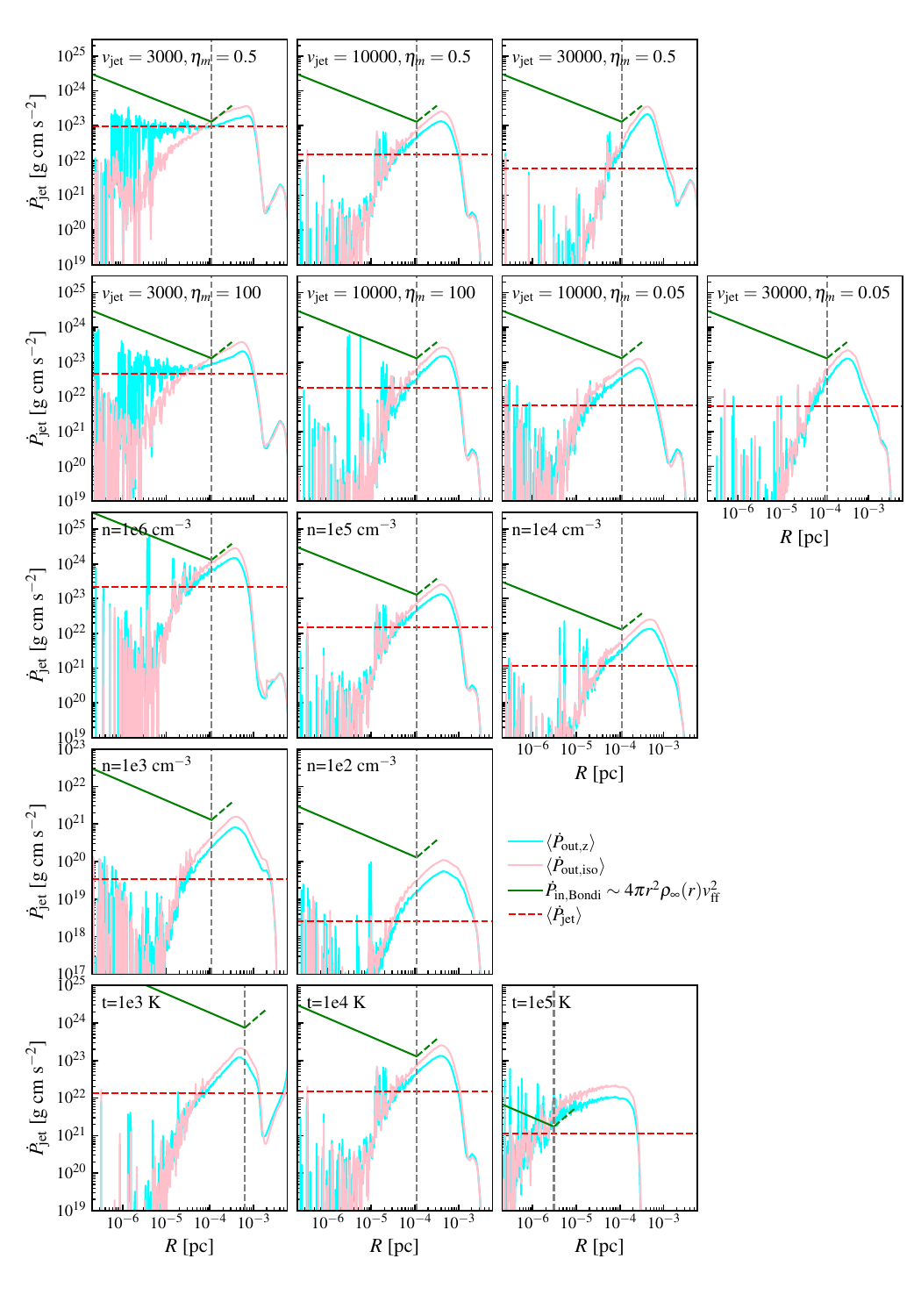}
\vspace{-1.2cm}
    \caption{The comparison of time-averaged momentum fluxes from the simulations is shown. Three types of momentum flux are illustrated: (i) the average jet momentum flux (red), (ii) the cocoon momentum flux, with cyan and pink lines representing the $z$ component and the isotropic component \ksu{(comparable to the lateral component)}, respectively, and (iii) the estimated inward free-fall (Bondi) momentum flux (green). The dashed gray vertical line in each plot indicates the Bondi radius. The isotropic component of the outward cocoon momentum flux matches the inward Bondi momentum flux at the Bondi radius. Runs with elongated cocoons (v=3000 km s$^{-1}$) have the $z$-component of their cocoon fluxes roughly matching the jet momentum fluxes (momentum-driven) and are much larger than the isotropic components. Runs with bubble-shaped cocoons (v$\gtrsim$10000 km s$^{-1}$, all except for the two panels with v=3000 km s$^{-1}$) all exhibit cocoon momentum fluxes (energy-driven) higher than the jet momentum fluxes.}
    \label{fig:balance_1e0_all}
\end{figure*}

\fref{fig:balance_1e0_all} shows that the regulation found in \citet{2023MNRAS.520.4258S} also occurs for the smaller black hole, $M_{\rm BH} = 1 {\rm M_\odot}$, which we explicitly tested in this work. In \fref{fig:balance_1e0_all}, we plot the injected jet momentum flux (red), the isotropic (pink) and z component (cyan) of the outflowing jet cocoon momentum flux, and the inflowing momentum flux assuming a Bondi solution (green) as a function of radius. Additionally, we indicate the Bondi radius with a vertical gray line.

In the low-jet-velocity ($V_{\rm jet} \sim 3000 \, {\rm km \, s}^{-1}$) cases, the jet cocoon clearly manifests momentum conservation in the z direction, as the z component of cocoon momentum flux (cyan) agrees with the injected jet momentum flux (red) to radii beyond the Bondi radius. The isotropic component of the cocoon momentum flux eventually picks up and becomes comparable to the z component of the cocoon momentum flux at the isotropization radius. Beyond this, the propagation of the jet cocoon becomes an energy-driven bubble, and the outflowing cocoon momentum flux can be higher than the injected momentum flux. Although at the Bondi radius the isotropic component of the cocoon momentum flux is much smaller than the z component momentum flux, it is the isotropic component that is regulated to the free-fall inflowing momentum flux at the Bondi radius. The overall picture is consistent with what is described in \sref{sec:self_regulation} for the case where $r_{\rm iso} > r_{\rm Bondi}$.

For runs with higher jet velocity, this isotropization happens at smaller radii, eventually occurring within the Bondi radius. However, in those cases, it is also the isotropic component of the cocoon momentum flux that matches the inflowing momentum flux, assuming the Bondi value at the Bondi radius. The overall picture is again consistent with what is described in  \sref{sec:self_regulation} for the case where $r_{\rm iso}<r_{\rm Bondi}$.

\subsubsection{Thermal phase structure of the cocoon/bubble gas}
\begin{figure}
    \centering
    \includegraphics[width=7.5cm]{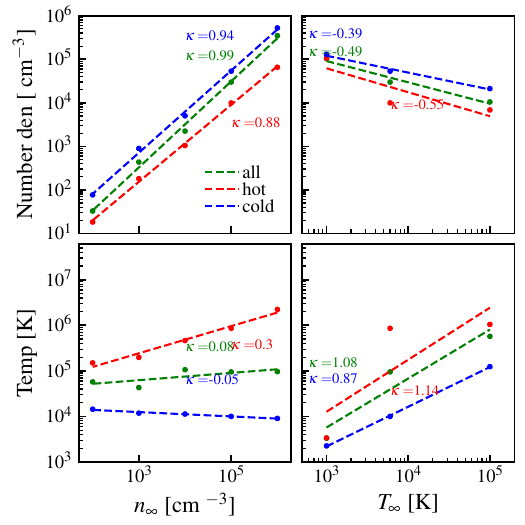}
    \caption{The dependence of the cocoon gas density and temperature on background gas properties (the former evaluated at the Bondi radius) for the $1 {\rm M_\odot}$ run is shown. The red, blue, and green dots and lines correspond to the hot, cold, and combined phases, respectively. The dots are from each simulation, and the lines are fitted power-laws with the index ($\kappa$) labeled. The cocoon is defined for simplicity as all gas with $T > 1.2 T_\infty$. We find $n_c \propto n_\infty^\zeta T_\infty^\xi$ with $\zeta \lesssim 0.9$ and $\xi \sim -0.6$ to $-0.7$.}
    \label{fig:phase_1e0}
\end{figure}
To determine how the jet mass flux scales with the gas properties and feedback parameters according to \Eqref{eq:mdotjet_reg}, we first need to understand how the thermodynamic properties of the cocoon gas scale with those of the background gas (parameterized as power-laws in \Eqref{eq:index}). Our previous study \citep{2023MNRAS.520.4258S} concluded that the scaling roughly follows $n_c \propto n_\infty^\zeta T_\infty^\xi$ with $\zeta \lesssim 0.9$ and $\xi \sim 0$ for $M_{\rm BH}=100 {\rm M_\odot}$.

As we explore a broader range of BH masses, \fref{fig:phase_1e0} describes the scaling for $M_{\rm BH} = 1 {\rm M_\odot}$. The y-axes represent the cocoon gas density and temperature, while the x-axes represent the background gas temperature and density. The red, blue, and green dots and lines correspond to the hot, cold, and combined phases, respectively. The dots are from each simulation, and the lines are fitted power laws with the index ($\kappa$) labeled. For simplicity, the cocoon is defined as all gas with $T > 1.2 T_\infty$. We find $n_c \propto n_\infty^\zeta T_\infty^\xi$ with $\zeta \lesssim 0.9$ and $\xi \sim -0.6$ to $-0.7$.

\subsubsection{The black hole accretion rate and jet mass flux} \label{sec:BH_jet_mass}
\begin{figure}
    \centering
    \includegraphics[width=7.5cm]{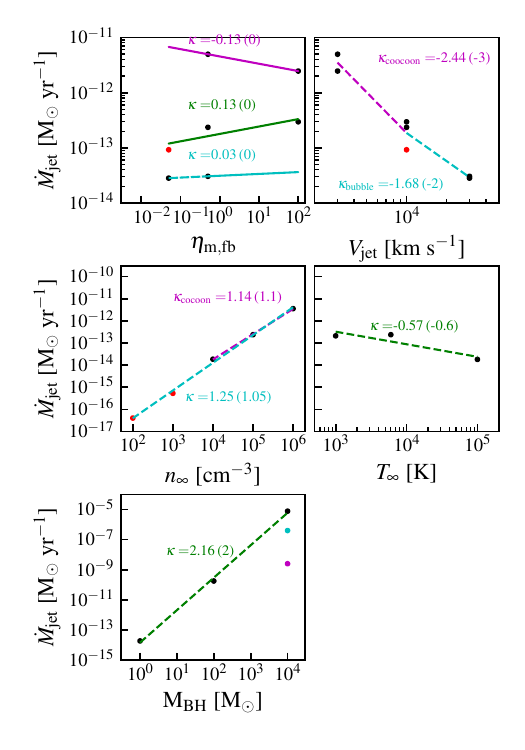}
    \caption{The dependence of the jet mass flux ($\dot{M}_{\rm jet}$) on the adopted jet model and background gas properties is shown for $M_{\rm BH} = 1 {\rm M_\odot}$ (top four panels) and across black hole masses assuming $(n_\infty, T_\infty, v_{\rm jet}) = (10^4 \, {\rm cm}^{-3}, 10^4 \, {\rm K}, 10^4 \, {\rm km \, s}^{-1})$ (bottom panel). The black, cyan, and magenta dots, respectively, characterize the accretion rate averaged over the first 0.5 Myr, all time, and after 1 Myr for $M_{\rm BH} = 10^4 {\rm M_\odot}$. The dots represent the simulation results, while the lines show power-law fits with the index ($\kappa$) labeled. The number in parentheses is an estimate from the toy model, and the fit to the cocoon gas-phase dependence in \fref{fig:phase_1e0} roughly agrees with what we measured from the simulation. The red dots represent runs that marginally failed to regulate, hence the slightly lower fluxes. In the last panel for the $10^4 {\rm M_\odot}$ case, the black, cyan, and magenta dots represent the accretion rate averaged over the first 0.5 Myr, the entire time, and after 1 Myr, respectively. The later decay in the jet fluxes due to core density suppression is discussed in \sref{sec:1e4}. }
    \label{fig:flux_v}
\end{figure}
With the scaling of gas properties relative to the background gas properties, we now have the full scaling of the black hole accretion rate with respect to the jet parameters and the gas properties (following \Eqref{eq:mdotjet_reg}). \fref{fig:flux_v} shows a comparison of what the toy model predicts versus what we measured in the simulation.

The top four panels in this figure show the results for a $1 {\rm M_\odot}$ black hole. The scaling obtained from the toy model (indicated in brackets) is broadly comparable to the fitted slope from the simulation.

We note that although the toy model predicts the black hole accretion rate assuming self-regulation at the Bondi radius, the required mass flux can be too high to achieve this. At any given time, the BH accretion rate and the wind mass flux can, at most, add up to the Bondi accretion flux. If the required fluxes exceed this, the accretion rate is capped at $\dot{M}_{\rm Bondi} \times (1 - \eta_{\rm jet, fb})^{-1}$, and the system simply fails to self-regulate. The red dots represent runs that marginally fail to regulate, which makes the fitted slope slightly different from the toy model results.

The bottom panel in \fref{fig:flux_v} shows the scaling of the black hole accretion rate as a function of the black hole mass, assuming $n_\infty = 10^4 \, {\rm cm}^{-3}$, $T_\infty = 10^4$ K, and $V_{\rm jet} = 10^4 , {\rm km \, s}^{-1}$. The resulting scaling is roughly $\dot{M}_{\rm jet} \sim M_{\rm BH}^2$, consistent with the prediction of the toy model.

We emphasize that we show three points for the $M_{\rm BH} = 10^4 {\rm M_\odot}$ case, corresponding to three different times. The black, cyan, and magenta dots, respectively, characterize the accretion rate averaged over the first 0.5 Myr, all time, and after 1 Myr. As time progresses, the accretion rate of the $10^4 {\rm M_\odot}$ run decays (see \fref{fig:flux_bh}). The lines were fitted only through the black dots (shortest averaging period). The resulting slope indicates that the initial BH accretion and self-regulation are well described by our toy model. Only at later times, when the density profile drops, does the result deviate from the model. We discuss this in the following section, \sref{sec:1e4}.

\subsection{Evolution of density profiles: beyond the $M_{\rm BH}=10^4\,{\rm M_\odot}$ threshold}\label{sec:1e4}
When the black hole mass reaches $M_{\rm BH} = 10^4 {\rm M_\odot}$, the jet cocoon can propagate beyond the core radius of approximately 1 pc, causing an overall decay of the density profile over time. In this section, we demonstrate this effect.

\subsubsection{The evolution of density profiles}
\begin{figure*}
    \centering
    \includegraphics[width=16cm]{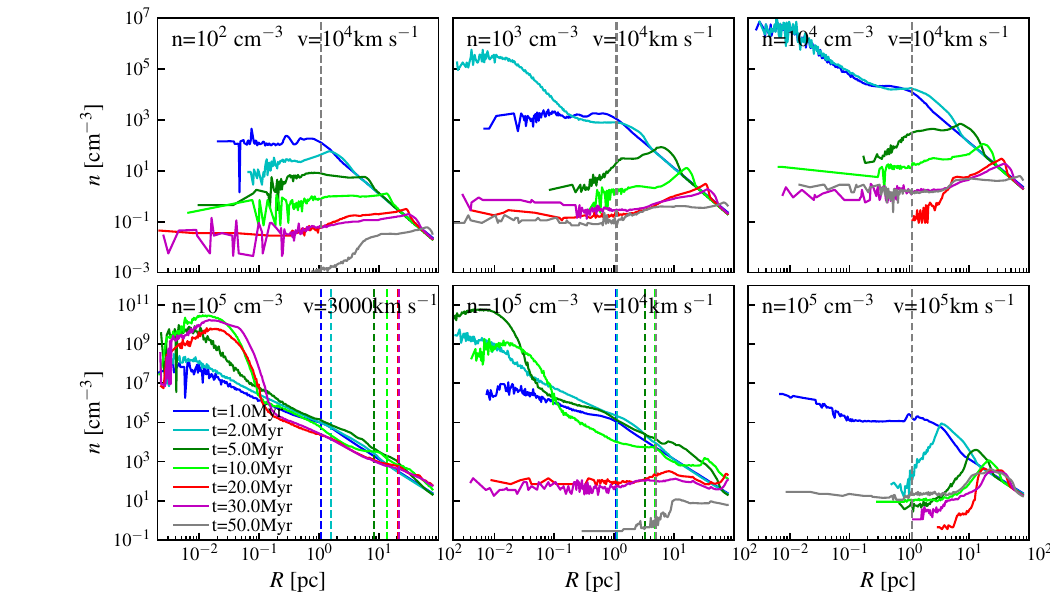}
    \caption{The time evolution of the density profile for the runs with $M_{\rm BH}=10^4 {\rm M_\odot}$ for a range of jet velocities and core densities (as indicated). For cases with very high jet velocity or a density lower than $10^4 \, {\rm cm}^{-3}$, the density within the radius where the cocoon propagates drops to at least the density at the jet cocoon shock front. For the case with $n_\infty=10^4 \, {\rm cm}^{-3}$ and $v_{\rm jet}=10^4 \, {\rm km \, s}^{-1}$, the density profile remains intact until the BH mass grows by a factor of 3, after which the density is suppressed. The density profile remains intact in the low jet velocity case (bottom left panel).}
    \label{fig:density_profile}
\end{figure*}
\fref{fig:density_profile}  shows the time evolution of the density profile for the runs with $M_{\rm BH}=10^4 {\rm M_\odot}$. For all such runs with a density lower than $10^4 \, {\rm cm}^{-3}$ and a velocity of $10^4 \, {\rm km \, s}^{-1}$ or above, the system experiences density drops over time. As the jet cocoon propagates, this density suppression extends to a larger radius. Eventually, the density suppression stops at approximately 30 pc, which is the terminal radius where the jet cocoon stops propagating (a model for this will be presented in \sref{sec:model_for_supression}). Within the terminal radius, the density is roughly uniform. The uniform density can be approximately explained by the high sound speed within the cocoon. \ksu{Assuming a density profile of $\rho_0 (r/r_0)^{-2}$, the resulting uniform density inside the shock front position, $R$, can be estimated as}
\begin{align}
\left<\rho\right>\sim \frac{\int\limits^R_0 4 \pi r^2 \rho_0 \left(\frac{r}{r_0}\right)^{-2} d r }{4\pi R^3/3}\sim 3\rho_0 
\left(\frac{R}{r_0}\right)^{-2}.
\end{align}
\ksu{Given that part of the mass is accreted by the black hole, this estimate is not far from the original density at the current position of the cocoon's shock front.}

The density suppression also depends on the jet velocity. For higher gas density at $10^5 \, {\rm cm}^{-3}$, we tested a wide range of jet velocities. For the fiducial jet velocity of $10^4 \, {\rm km \, s}^{-1}$, the density profile is initially sustained at the initial profile for about 100 kyr, during which the BH mass grows by a factor of 3. After that, density suppression starts to occur, and the terminal radius reaches roughly 100 pc. For the lower jet velocity case, density suppression never occurs. Instead, the density evolves from an initially flat profile within the core radius toward a cuspier profile. For the higher jet velocity case, density suppression occurs again.

Overall, the lower the gas density and/or the higher the jet velocity, the more likely density suppression is to happen. The overall trend suggests that the more bubble-like the jet cocoon is, the easier it is for density suppression to occur. We discuss the criteria for this as follows.

\subsubsection{Criteria for density suppression}
\begin{figure*}
    \centering
    \includegraphics[width=16cm]{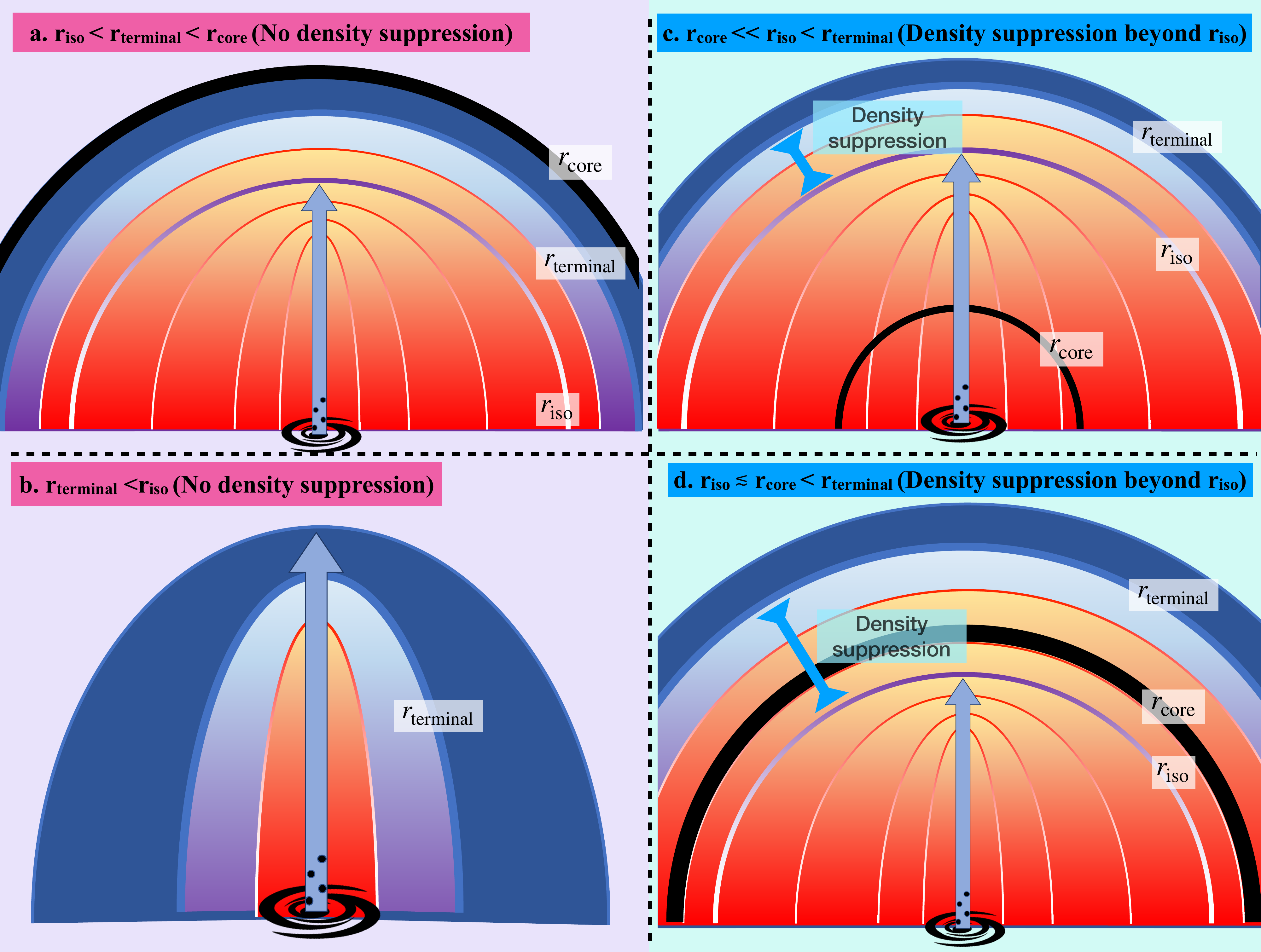}
    \caption{A cartoon depicting the conditions under which an overall drop in the density profile occurs. These conditions are determined by comparing the core radius ($r_{\rm core}$), within which the density is initially flat; the isotropization radius ($r_{\rm iso}$), where the jet cocoon becomes isotropic; and the terminal radius ($r_{\rm terminal}$), where the jet cocoon or bubble loses its energy and stalls.}
    \label{fig:cartoon}
\end{figure*}

To understand the conditions where density suppression occurs, we highlight three relevant radii: the core radius ($r_{\rm core}$), within which the density is initially flat; the isotropization radius ($r_{\rm iso}$), where the jet cocoon becomes isotropic; and the terminal radius ($r_{\rm terminal}$), where the jet cocoon or bubble stops propagating. Details of the terminal radius are discussed in  \sref{sec:model_for_supression}. By comparing these three radii, we can identify several different regimes, as characterized in the cartoon shown in \fref{fig:cartoon}. The fourth relevant radius is the Bondi radius ($r_{\rm Bondi}$). If $r_{\rm Bondi}$ falls within the region of suppressed density, the black hole 'feels' the density suppression, and the black hole accretion rate is reduced.  We list them case by case as follows:

\begin{enumerate}[(a)]
\item{\bf $r_{\rm iso}<r_{\rm terminal}<r_{\rm core}$ (No density suppression):} As shown in the upper left of  \fref{fig:cartoon}, the cocoon stops propagating before reaching the core radius, so it never reaches the radius where the density starts to drop. Therefore, there is no density suppression. This is typical for smaller black hole cases ($M_{\rm BH}<10^4 {\rm M_\odot}$).
\item{\bf $r_{\rm terminal}<r_{\rm iso}$ (No density suppression):} As shown in the lower left of  \fref{fig:cartoon}, the jet cocoon never isotropizes before losing all its energy. As a result, even if the cocoon penetrates into the low-density region, it does not suppress the density in all directions. Overall, the spherically averaged density profile is not significantly suppressed.
\item{\bf $r_{\rm core}<r_{\rm iso}<r_{\rm terminal}$ (Density suppression beyond $r_{\rm iso}$):} As shown in the upper right of  \fref{fig:cartoon}, the jet cocoon isotropizes beyond the core radius and continues to propagate into the low-density region. The density then becomes constant from the isotropization radius up to the radius where the jet cocoon shock is. This is typical for more massive black holes ($M_{\rm BH}>10^4 {\rm M_\odot}$) with isotropization radius larger than the core radius.
\item{\bf $r_{\rm iso}<r_{\rm core}<r_{\rm terminal}$ (Density suppression beyond $r_{\rm iso}$):} As shown in the lower right of  \fref{fig:cartoon}, the jet cocoon first isotropizes and then pushes beyond the core radius into the low-density region. As indicated in  \fref{fig:density_profile}, the density then becomes constant from the isotropization radius up to the radius of the jet cocoon shock. This is typical for more massive black holes ($M_{\rm BH}>10^4 {\rm M_\odot}$) with isotropization radius smaller than the core radius.
\end{enumerate}

\subsubsection{The consequence of density suppression -- a secularly evolving regulation}

\begin{figure*}
    \centering
    \includegraphics[width=16cm]{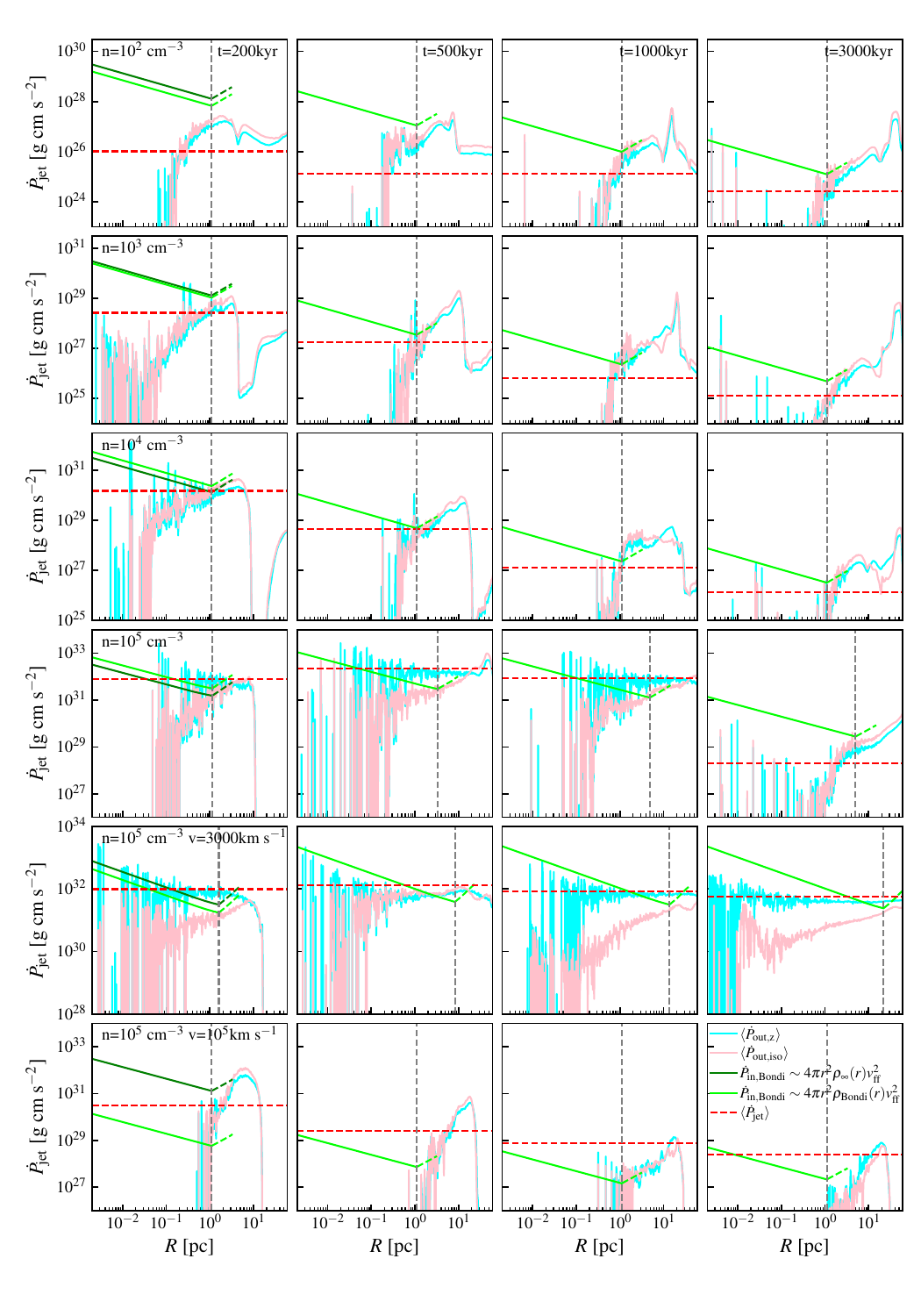}
    \vspace{-1cm}
    \caption{The comparison of time-averaged momentum fluxes from six simulations with $M_{\rm BH} = 10^4 {\rm M_\odot}$. Each row represents one run, with each column corresponding to a different time of that run. Three types of momentum flux are illustrated: (i) the average jet momentum flux (red), (ii) the cocoon momentum flux, with cyan and pink lines representing the $z$ component and the isotropic component \ksu{(comparable to the lateral component)}, respectively, and (iii) the estimated inward Bondi momentum flux assuming the initial density (green) and real-time density (lime). The dashed gray vertical line in each plot indicates the Bondi radius. The isotropic component of the outward cocoon momentum flux broadly matches the instantaneous (lime) inward Bondi momentum flux at the Bondi radius, tracking the decrease in density at the Bondi radius.}
    \label{fig:balance_1e4}
\end{figure*}

As the density profile drops, the density at the Bondi radius also decreases. Meanwhile, the accretion rate and jet fluxes also drop. However, the isotropic component of the outflowing cocoon momentum flux still regulates to the real-time Bondi inflowing momentum flux. The only difference is that it becomes a moving regulation.

\fref{fig:balance_1e4} demonstrates this moving regulation for six different simulations with a $10^4 {\rm M_\odot}$ BH. As in \fref{fig:balance_1e0_all}, we plot the injected jet momentum flux (red), the isotropic (pink) and z component (cyan) of the outflowing jet cocoon momentum flux, and the inflowing momentum flux assuming a Bondi solution based on the initial density (green) as a function of radius. To account for the decay of density at the Bondi radius, we also plotted the Bondi inflowing momentum flux corrected by the real-time measured density at the Bondi radius (lime). We indicate the evolving Bondi radius at each time with a vertical gray line.

Each row represents one run at different times. For all the runs, the real-time corrected Bondi inflowing momentum flux matches the isotropic component of the cocoon momentum flux at the Bondi radius, indicating that this moving regulation holds in a broad sense. We note that the run with $n=10^5 \, {\rm cm}^{-3}$, $v=3000\, {\rm km \, s}^{-1}$ (5th row) never experiences density suppression as time progresses and the black hole mass increases. The regulation still holds perfectly.

\begin{figure*}
    \centering
    \includegraphics[width=16cm]{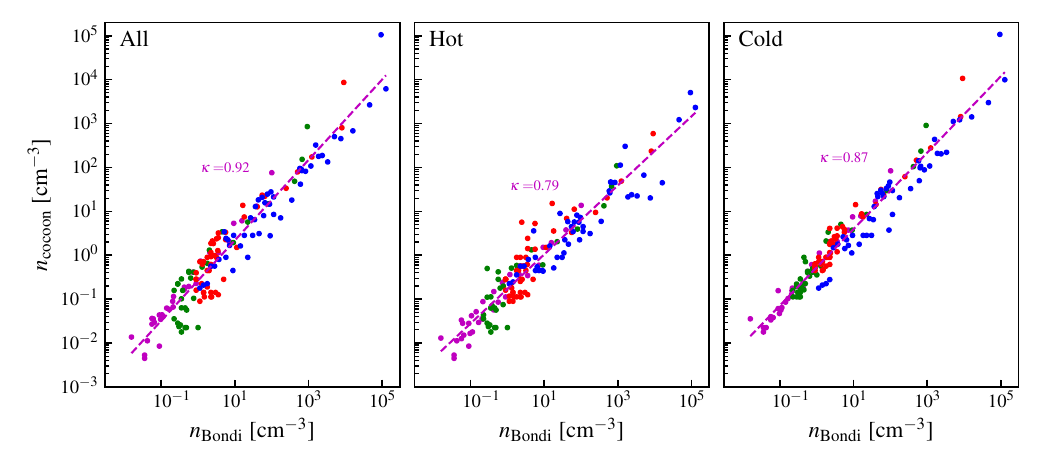}
    \caption{The dependence of the cocoon gas density on background gas properties (evaluated at the Bondi radius) for the $10^4 {\rm M_\odot}$ run. Each simulation is divided into 0.1 Myr periods, and the average quantities for each period are plotted. The dots of the same color are from the same simulation. The lines are fitted power laws with the index ($\kappa$) labeled. The cocoon is defined for simplicity as all gas with $T > 1.2 T_\infty$, and hot gas as gas with $T > 3.6 T_\infty$. We find $n_c \propto n_\infty^\zeta T_\infty^\xi$ with $\zeta \lesssim 0.9$.}
    \label{fig:den_temp_fit_1e4}
\end{figure*}

As the density at the Bondi radius drops, the gas density within the jet cocoon also decreases. Therefore, we need to account for this when determining the scaling relation of cocoon gas density with respect to background gas density (as shown in  \fref{fig:phase_1e0}) for the $M_{\rm BH}=10^4 {\rm M_\odot}$ case. In \fref{fig:den_temp_fit_1e4}, we divide the simulation into several 0.1 Myr bins and plot the average background gas density at the Bondi radius ($n_{\rm Bondi}$) and the cocoon gas density at the Bondi radius ($n_{\rm cocoon}$) at the corresponding times. As a result of the time-varying regulation, each run forms a series of points, and we fit a line through all the points. We find that $n_{\rm cocoon} \propto n_{\rm Bondi}^\zeta$ with $\zeta \lesssim 0.9$, very similar to the scaling relation obtained for other black hole masses without the overall density evolution.

\begin{figure}
    \centering
    \includegraphics[width=8cm]{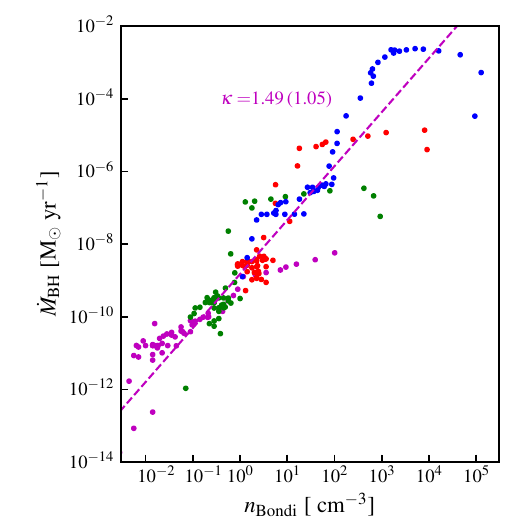}
    \caption{The dependence of the jet mass flux ($\dot{M}_{\rm jet}$) on the background gas properties for $M_{\rm BH}=10^4 {\rm M_\odot}$. The run is divided into 0.8 Myr periods, and the accretion rate and background gas density at the Bondi radius averaged over each specific time period are plotted. Each color represents one run. The lines show power-law fits, with the index ($\kappa$) labeled. The number in parentheses is an estimate from the toy model, and the fit to the cocoon gas-phase dependence in \fref{fig:den_temp_fit_1e4} roughly agrees with what we measured from the simulation. }
    \label{fig:mdot_fit_1e4}
\end{figure}
Finally, to demonstrate that the toy model picture works even for the moving regulation case, we can plug the scaling relation (between the cocoon gas properties and the background gas properties) into the toy model to obtain the scaling of the black hole accretion rate with the background gas properties. \fref{fig:mdot_fit_1e4} again divides each simulation into 0.8 Myr time periods and calculates the $\dot{M}_{\rm BH}$ and $n_{\rm cocoon}$ averaged over the specific time period. By fitting through all the points, we obtain a scaling relation $\dot{M}_{\rm BH} \propto n_{\rm Bondi}^{1.5}$, broadly similar to (but slightly steeper than) what the toy model implies. The scaling relation is also broadly consistent with what was obtained for the $M_{\rm BH} = 1 {\rm M_\odot}$ and $M_{\rm BH} = 10^4 {\rm M_\odot}$ cases.

\subsection{Summary for the toy model for self-regulation as a function of BH mass}\label{sec:summary_model}
\begin{figure*}
    \centering
    \includegraphics[width=16cm]{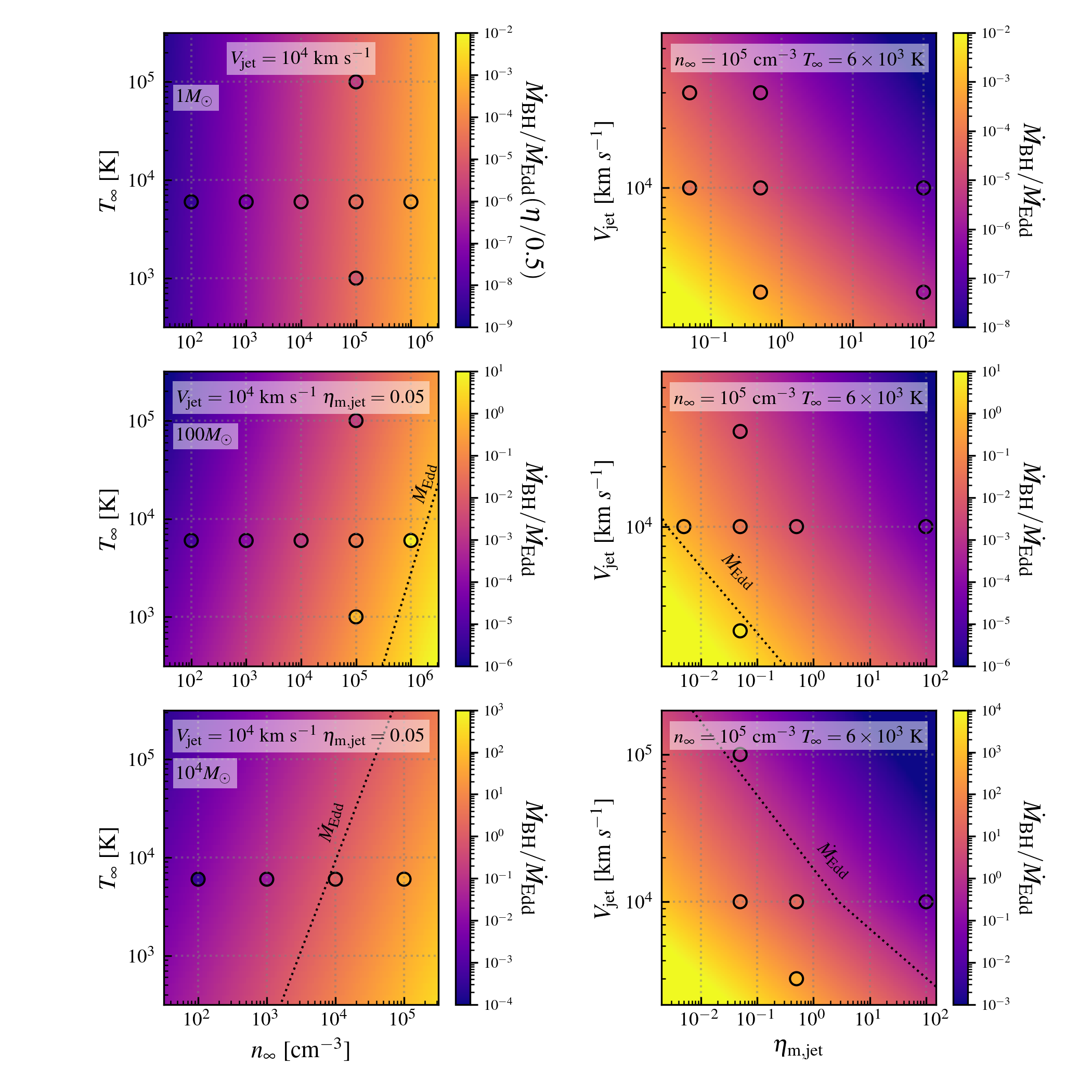}
    \caption{
The predicted $\dot{M}_{\rm acc}/\dot{M}_{\rm Edd}$ from the scaling of our toy model, assuming normalization to the fiducial runs, is shown as the background color in each panel. Runs with low background gas temperature ($T_\infty$), high background gas density ($n_\infty$), low jet velocity ($V_{\rm jet}$), or low feedback mass loading ($\eta_{\rm m, jet}$) result in super-Eddington accretion for cases with $M_{\rm BH} \gtrsim 100 {\rm M_\odot}$. Black dotted lines indicate Eddington accretion. The results from the simulations are shown as circles, colored with the measured value in order to give an indication of how well the model fares in predicting the simulation results.
The three rows show the cases for $1 {\rm M_\odot}$, $100 {\rm M_\odot}$, and the first 0.2 Myr of $10^4 {\rm M_\odot}$. The left column shows the results as a function of gas properties, while the right column shows the results as a function of AGN jet parameters. They show a qualitative agreement with the toy model.  }
    \label{fig:mdot_map}
\end{figure*}
Here we summarize what our generalized analytic model predicts for the scalings, and compare it with the simulation results. Compiling the scaling relations we derived, we have:

\begin{align}\label{eq:mdotbh}
\frac{\dot{M}_{\rm BH}}{\dot{M}_{\rm Edd}}\sim \begin{cases}
     \propto  M_{\rm BH} \rho_\infty^{\alpha_1} \,\,T_\infty^{\tau_1}\,\eta_{\rm m, BH}^{-1} V_{\rm jet}^{-3}  \text{\,\,\,\,(for\,\,} r_{\rm iso}>r_{\rm Bondi})\\
     \,\\
      \propto M_{\rm BH} \rho_\infty^{\alpha_2}T_\infty^{\tau_2} \eta_{\rm m, BH}^{-1} V_{\rm jet}^{-2}  \text{\,\,\,\,\,(for\,\,} r_{\rm iso}<r_{\rm Bondi})
    \end{cases} 
\end{align}
where 
\begin{align}\label{eq:mdotbh_power}
&\alpha_1\sim 1.1 \,\,\,\,\alpha_2\sim1.05\notag\\
&\tau_1\sim 0.6 \text{ (for $M_{\rm BH}=1{\rm M_\odot}$) to } -0.2 \text{ (for $M_{\rm BH}=100\,{\rm M_\odot}$)}\notag\\
&\tau_2\sim -0.2 \text{ (for $M_{\rm BH}=1{\rm M_\odot}$) to } -0.6 \text{ (for $M_{\rm BH}=100\,{\rm M_\odot}$)}
\end{align}
Note that from the simulations, we measure $\alpha_1 \sim \alpha_2 \sim 1.5$.

\fref{fig:mdot_map} shows the comparison of the scaling from the toy model to the simulation results. The background color represents the toy model's predictions, while each circle corresponds to one run, colored by the actual measured accretion rate from the simulation. The first three rows show the results for $1 {\rm M_\odot}$, $100 {\rm M_\odot}$, and the first 0.2 Myr of $10^4 {\rm M_\odot}$, respectively. For the left panel, which shows the scaling of the accretion rate with gas temperature and density, we assume $r_{\rm iso} > r_{\rm Bondi}$, as it describes most of the parameter space there. All of them show reasonable agreement with the toy model's prediction. We note that for BH masses larger than $100 {\rm M_\odot}$, super-Eddington accretion is possible in part of the parameter space with low feedback efficiency, high gas density, and low gas temperature.

For $M_{\rm BH} = 10^4 {\rm M_\odot}$, there is a significant decay in the accretion rate after the initial period, as shown in \fref{fig:flux_bh}.  We provided an analytical description of how the jet cocoon propagates once density suppression occurs in \sref{sec:model_for_supression}.

\section{An Analytic Model for the Terminal Radius of the Jet Cocoon Expansion}\label{sec:model_for_supression}

In this section, we generalize the model previously developed to provide a quantitative understanding of how the jet cocoon evolves once density suppression occurs. From the discussion above, we know the following facts when the density is suppressed:
\begin{itemize}
\item The isotropic component of the cocoon momentum flux is regulated to the real-time inflowing Bondi momentum flux, given the current density at the Bondi radius. Thus, $\dot{M}_{\rm BH} \propto n_{\rm Bondi}^\alpha$.
\item Density suppression occurs beyond $r_{\rm iso}$ when $r_{\rm terminal} > r_{\rm iso}$ and $r_{\rm terminal} > r_{\rm core}$. 

\item \ksu{The black hole can `'feel'' the density suppression if the Bondi radius, $r_{\rm Bondi}$, is larger than $r_{\rm iso}$ but smaller than $r_{\rm terminal}$, as the density at $r_{\rm Bondi}$ will decrease over most of the solid angle. For simplicity, we do not consider density suppression when $r_{\rm Bondi} < r_{\rm core}$.}
\item \ksu{Density suppression begins to affect the black hole as soon as the jet cocoon propagates through the Bondi radius.}
\item When density suppression occurs, the density within the suppressed region becomes roughly constant at the density of the current position of the cocoon shock front. 
\end{itemize}
The key assumption we will make to derive the location of the terminal radius is that the jet cocoon terminates when the cumulative energy flux of the jet equals the time-integrated cooling rate within the jet cocoon. Once this occurs, the jet cocoon can no longer propagate, and
the influence of the cocoon on the surrounding gas becomes minimal. Equipped with this information, we can write down an analytical description of cocoon propagation.
 In \sref{sec:jet_e_flux}, we will derive the integrated jet energy flux, in \sref{sec:e_cooling}, we will compute the cocoon radiative cooling rate and, in \sref{sec:terminal_r}, we will equate these to determine the terminal radius.

\subsection{Integrated jet energy flux} \label{sec:jet_e_flux}

\ksu{Given the points above, we consider density suppression to start the moment ($t=t_{\rm Bondi}$) when an isotropized jet cocoon, with radius $R(t)$, passes through the Bondi radius, which is equal to or larger than the core radius. We will use $\rho_{\rm 0,\,Bondi}$ ($n_{\rm 0,\,Bondi}$) to denote the initial background gas (number) density at the Bondi radius, $r_{\rm Bondi}$ when $t=t_{\rm Bondi}$. We note that for our runs with an initial black hole mass of $10^4\,{\rm M_\odot}$, this radius is $\sim1$ pc, which is also the core radius, $r_{\rm core}$. In that case, the initial number density at the Bondi radius, $n_{\rm 0,\,Bondi}$, is also the initial core gas density, $n_{\rm 0,\,core}$.} 

\ksu{In our previous paper \citep{2023MNRAS.520.4258S}, we derived the energy conservation equation governing the conversion of jet energy flux into cocoon energy flux during the bubble ($R(r)>r_{\rm iso}$) phase, as well as the momentum conservation at the cocoon shock front for a uniform-density background medium. Given the density evolution within the Bondi radius, the gas density also becomes a function of time.}
\begin{align}\label{eq:moving_eng_conservation}
4\pi R(t)^2 \rho_c(t) V_{\rm R,\, Hot}^3(t) &=\frac{\gamma }{2} \dot{M}_{\rm jet}(t) V_{\rm jet}^2\notag\\
V_{\rm R,\, Hot}^2(t) \rho_c(t)&=V_{\rm R}^2(t) \rho(t),
\end{align}
\ksu{where  $V_{\rm R,\,Hot}(t)$ and $\rho_c(t)$ are the velocity and density of the hot phase of gas, respectively, while $V_{\rm R}(t)$ is the cocoon propagation velocity, and $\rho(t)$ is the density at the shock front of the jet cocoon.}

When $R(t)$ reaches $r_{\rm Bondi}$ at $t=t_{\rm Bondi}$, we define:
\begin{align}
\rho(t_{\rm Bondi})=\rho_{\rm 0,\,Bondi},\,\,\, \rho_c(t_{\rm Bondi})=\rho_{\rm 0,\,c}, \notag\\
{\rm and}\,\,\, \dot{M}_{\rm jet}(t_{\rm Bondi})=\dot{M}_{\rm 0,\, jet}. 
\end{align}
\ksu{After $R(t)$ reaches $r_{\rm Bondi}$, the gas between $r_{\rm iso}$ and $R(t)$ remains at approximately constant density. Since $r_{\rm Bondi}$ lies within this radial range, $\rho_{\rm Bondi}(t) = \rho(t)$.}

\ksu{Since the density profile follows $\rho(r) \propto n(r) \propto r^{-2}$ in the radial range we are considering, we have}
\begin{align}\label{eq:rho_scaling_r}
\rho_{\rm Bondi}(t)&=\rho(t)=\rho_{\rm 0,\,Bondi}\left(\frac{R(t)}{r_{\rm 0,\, Bondi}}\right)^{-2} \notag\\
\rho_{\rm c}(t)&=\rho_{\rm 0,\,c}\left(\frac{R(t)}{R_{\rm core}}\right)^{-2}.
\end{align}

Supposed $\dot{M}_{\rm jet} (t)\propto \rho_{\rm Bondi}^\alpha (t)\propto n_{\rm Bondi}^\alpha (t)$, we have
\begin{align}\label{eq:mdot_scaling_r}
\dot{M}_{\rm jet}(t)=\dot{M}_{\rm 0,\,jet} \left(\frac{\rho(t)}{\rho_{\rm 0,\,Bondi}}\right)^\alpha=\dot{M}_{\rm 0,\,jet} \left(\frac{R(t)}{r_{\rm Bondi}}\right)^{-2 \alpha}.
\end{align}
\ksu{Following the toy model summarized in \sref{sec:jet_propogation_constant}, \Eqref{eq:index} can be generalized as $n_c(t) \propto n_{\rm Bondi}^\xi(t)$ with $\xi \lesssim 1$.  \Eqref{eq:mdotjet_reg} also gives $\alpha \sim (3 - \xi)/2$ for the isotropic case.}

Putting \Eqref{eq:moving_eng_conservation} \Eqref{eq:rho_scaling_r}, and \Eqref{eq:mdot_scaling_r} together, we get:
\begin{align}
V_R(R)=\left(\frac{\gamma}{8\pi}\right)^{1/3} \dot{M}_{\rm0,\,jet}^{1/3}V_{\rm jet}^{2/3} \rho_{0,\,c}^{1/6} \rho_{\rm 0,\,Bondi}^{-1/2}r_{\rm Bondi}^{\frac{-2+2\alpha}{3}}R^{-2\alpha/3}(t),
\end{align}
and 
\begin{align}
R(t)=&\left[\left(\frac{2\alpha+3}{3}\right)^3\frac{\gamma}{8\pi}\right]^{\frac{1}{3+2\alpha}} \dot{M}_{\rm 0,\,jet}^{\frac{1}{3+2\alpha}} V_{\rm jet}^{\frac{2}{3+2\alpha}}\rho_{0,\,c}^{\frac{1}{6+4\alpha}}\rho_{\rm 0,\,Bondi}^{\frac{-3}{6+4\alpha}}\notag\\
& r_{\rm Bondi}^{\frac{-2+2\alpha}{3+2\alpha}} t^{\frac{3}{2\alpha+3}}.
\end{align}

The net integrated energy flux from the time when $R(t_{\rm Bondi})=r_{\rm Bondi}$ is
\begin{align}
{E}_{\rm tot,\, jet}(t) =&\int \frac{1}{2} \dot{M}_{\rm jet} V_{\rm jet}^2 dt\notag\\
                        =&\int\limits_{t_{\rm Bondi}}^t  \frac{1}{2} \dot{M}_{\rm 0,\,jet} \left(\frac{R(t)}{R_{\rm Bondi}}\right)^{-2 \alpha} V_{\rm jet}^2  dt\notag\\
                        =&\frac{1}{2} \left[\left(\frac{2\alpha+3}{3}\right)^3\frac{\gamma}{8\pi}\right]^{\frac{-2\alpha}{3+2\alpha}} \dot{M}_{\rm 0,\,jet}^{\frac{3}{3+2\alpha}} V_{\rm jet}^{\frac{6}{3+2\alpha}}\rho_{\rm 0,\,c}^{\frac{-2\alpha}{6+4\alpha}}\rho_{\rm 0,\,Bondi}^{\frac{6\alpha}{6+4\alpha}}\notag\\
& r_{\rm Bondi}^{\frac{10\alpha}{3+2\alpha}} \left[ \frac{2\alpha+3}{4\alpha-3}\left(t_{\rm Bondi}^{\frac{3-4\alpha}{2\alpha+3}}-t^{\frac{3-4\alpha}{2\alpha+3}}\right)\right],
\end{align}
or in terms of the position of the shock front,

\begin{align}\label{eq:edot_t_after_sup}
{E}_{\rm tot, J}(R) =&\int \frac{1}{2} \dot{M}_{\rm jet} V_{\rm jet}^2 dt\notag\\
                        =&\int\limits_{r_{\rm Bondi}}^R \frac{1}{2} \dot{M}_{\rm0,\,jet} \left(\frac{R}{r_{\rm Bondi}}\right)^{-2 \alpha} V_{\rm jet}^2 V_R^{-1}(R) dR\notag\\
                        =&\left(\frac{\gamma}{\pi}\right)^{-1/3}\dot{M}_{\rm0,\,jet}^{2/3}r_{\rm Bondi}^{\frac{2+4\alpha}{3}} V_{\rm jet}^{4/3} \rho_{\rm0,\,c}^{1/6} \rho_{\rm 0,\,Bondi}^{1/2} \notag\\
                        &\left(\frac{3}{4\alpha-3}\right)\left(r_{\rm Bondi}^{\frac{3-4\alpha}{3}} -R^{\frac{3-4\alpha}{3}} \right).
\end{align}

This saturates to a maximum value:
\begin{align}
{E}_{\rm max,\,jet} &=\left(\frac{\gamma}{\pi}\right)^{-1/3}\dot{M}_{\rm0,\,jet}^{2/3}r_{\rm Bondi}^{\frac{5}{3}} V_{\rm jet}^{4/3} \rho_{\rm 0,\,c}^{-1/6} \rho_{\rm 0,\, Bondi}^{1/2} \left(\frac{3}{4\alpha-3}\right)\notag\\
&\propto n_{\rm 0,\,Bondi}^{\frac{2\alpha}{3}-\frac{\xi}{6}+\frac{1}{2}}.
\end{align}
\ksu{From \fref{fig:den_temp_fit_1e4} and \fref{fig:mdot_fit_1e4}, where we measured $\xi = 0.9$ and $\alpha = 1.5$, we obtain ${E}_{\rm max,\,jet}\propto n_{\rm 0,\,Bondi}^{1.35}$} \footnote{\ksu{Alternatively, if we only use $\xi = 0.9$ from \fref{fig:den_temp_fit_1e4}  and substitute it into  \Eqref{eq:mdotjet_reg} to get  $\alpha=1.05$, we find ${E}_{\rm max,\,jet}\propto n_{\rm 0,\, Bondi}^{1.05}$.}}.

\subsection{The integrated cocoon cooling} \label{sec:e_cooling}
The instantaneous cooling rate within the jet cocoon at a specific time after $R(t_{\rm Bondi})=r_{\rm Bondi}$ is given by:
\begin{align}
\dot{E}_{\rm cool} (R) &=\int\limits^R_0\frac{d \dot{E}_{\rm cool}}{d\,{\rm Vol}} 4\pi r^2 dr\notag\\
&=\int\limits^R_0 \Lambda 4\pi  n_{\rm c,\,e}(R) n_{\rm c,\,H}(R) r^2 dr\notag\\
                        &\sim \Lambda \frac{4\pi}{3} n_{\rm c,\,e}(R)n_{\rm c,\,H}(R) R^3 \notag\\
                        &\sim \Lambda' \frac{4\pi}{3} n_{\rm 0,\,c}^2 \left(\frac{R}{r_{\rm Bondi}}\right)^{-4} R^3, 
\end{align}
\ksu{where $\Lambda$ is the cooling function, and $n_{\rm c,\,e}(R)$ and $n_{\rm c,\,H}(R)$ denote the electron and hydrogen number densities of the jet cocoon gas, respectively. Given the assumption of constant density within the jet cocoon, both are constant with respect to $r$ but scale with $R$, the current position of the cocoon shock front, following \Eqref{eq:rho_scaling_r}. $\Lambda'$ accounts for the conversion from hydrogen or electron number density to the overall gas cocoon number density.} To estimate the cooling function, the cocoon gas is mostly from $10^5 - 10^8$ K, for which $\Lambda \sim 10^{-23} {\rm erg\, s}^{-1} {\rm cm}^{3}$.

The accumulated cooling up $R(t_{\rm Bondi})=r_{\rm Bondi}$ to the time the bubble reaches $R$ would then be:
\begin{align}
E_{\rm cool} (R)= &\int \dot{E}_{\rm cool} (t)  dt\notag\\
                =&\int\limits^{R}_{r_{\rm Bondi}} \dot{E}_{\rm cool}(R) V_R^{-1}(R) dR\notag\\
                = &\frac{4\pi}{3} \Lambda' n_{\rm 0,\,c}^2  r_{\rm Bondi}^{\frac{-2\alpha+14}{3}} \left(\frac{\gamma}{8\pi}\right)^{-1/3}\dot{M}_{\rm 0,\,jet}^{-1/3}\notag\\
                & V_{\rm jet}^{-2/3} \rho_{\rm 0,\,c}^{-1/6}\rho_{\rm 0,\,Bondi}^{1/2}\left(\frac{3}{2\alpha}\right)\left(R^{\frac{2\alpha}{3}}-r_{\rm Bondi}^{\frac{2\alpha}{3}}\right)\notag\\
                \propto & \, n_{\rm0,\,c} ^{11/6} n_{\rm0,\,Bondi} ^{1/2-\alpha/3}  R^{\frac{2\alpha}{3}}\,\,\,\,\,\,\,\,\,\,({\rm for}\,\, R>>r_{\rm Bondi})
\end{align}

\subsection{The terminal radius}\label{sec:terminal_r}
Finally, we derive the terminal radius by balancing heating and cooling, or in other words setting $\dot{E}_{\rm cool} (R)\sim \dot{E}_{\rm max,\,jet} (R)$ and solving for $R=r{\rm terminal}$. For $\alpha=1.5$ and $\xi=0.9$, we have
\begin{equation}
r_{\rm terminal}\sim 12 \, {\rm pc} \left(\frac{n_{\rm0,\,Bondi}}{10^3\,{\rm cm}^{-3}}\right)^{0.3}\sim 12 \, {\rm pc} \left(\frac{n_{\rm0,\,core}}{10^3\,{\rm cm}^{-3}}\right)^{0.3}.
\end{equation}
\ksu{We find a very weak dependence on the gas density, which is roughly comparable to the value shown in \fref{fig:density_profile}.}

\section{Putting it All Together: Predicting seed BH growth for an atomic-cooling halo}\label{sec:predict}

In this section, we combine the analytic model we have developed with the expected conditions for atomic-cooling halos to determine how much BH growth we expect under various conditions.  In \sref{s:phase_of_growth}, we discuss the detailed calculations behind this plot and the different phases of growth. In \sref{s:seed}, we explore how these results change with BH seed mass and time, and in \sref{sec:maximum_feedback_efficiency}, we highlight the importance of the feedback efficiency.  We first review the contents of our analytic understanding, which includes:
\begin{itemize}
\item How a jet cocoon propagates in a constant density environment (\sref{sec:jet_propogation_constant}).
\item How a jet cocoon propagates in an $n \propto r^{-2}$ environment undergoing density suppression (\sref{sec:model_for_supression})
\item How black hole accretion is regulated by jets in a constant density environment  (\sref{sec:self_regulation}).
\item How black hole accretion is regulated by jets in an $n \propto r^{-2}$ decay environment undergoing density suppression (\sref{sec:terminal_r}).
\end{itemize}
Equipped with this knowledge, we can predict the growth of a black hole seeded in the fiducial density profile, with $n=10^5\, {\rm cm}^{-3}$ within $r_{\rm core}=1$ pc and an $n \propto r^{-2}$ profile outside of that, typical of high-redshift ($z \sim$ 20) atomic-cooling halos \citep[e.g.,][]{2019MNRAS.486.3892R}. 

\begin{figure}
    \centering
    \includegraphics[width=8.5cm]{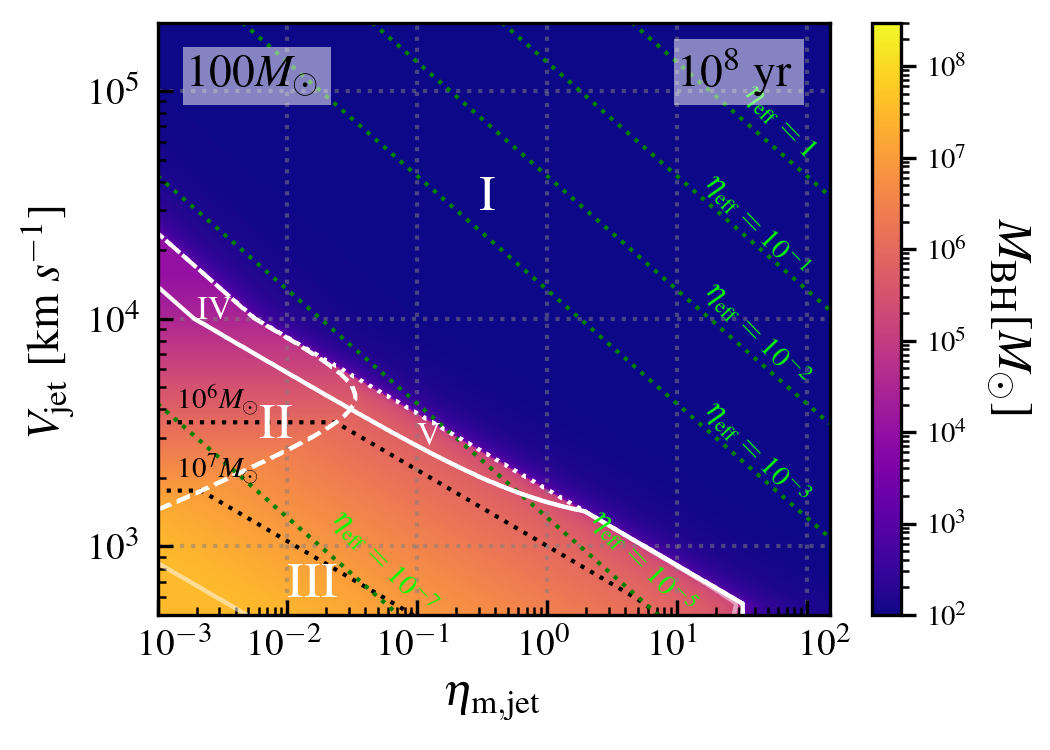}
    \caption{The prediction of the black hole mass 100 Myr after seeding a 100 ${\rm M_\odot}$ black hole in our fiducial profile of an atomic-cooling halo is as follows: Black holes with high-efficiency feedback (I, above the dotted line) never grow out of the constant-density phase. Black holes with low-efficiency feedback (II \& III, below the solid white line) undergo an extended ``fail to regulate'' phase, where most of the mass is accreted. \ksu{If the feedback efficiency is very low (below the light white line), the black hole has not reentered the self-regulation phase at the time.} Black holes with intermediate-efficiency feedback (IV \& V, between the dashed and solid white lines) never go through a ``fail to regulate'' phase. \ksu{Black holes with feedback parameters to the left of the dashed line (II \& IV) end up going through a core density suppression phase.} The maximum efficiency allowed for forming a supermassive black hole with $M_{\rm BH}>10^6 {\rm M_\odot}$ is $\eta_{\rm eff}=10^{-5}$. See \sref{s:phase_of_growth} for discussion.}
    \label{fig:predict_m_100}
\end{figure}

For a seed black hole mass of $100 \, {\rm M_\odot}$ (an expected outcome of first star formation), the predicted black hole mass after $10^8$ years as a function of the AGN feedback parameter is shown in \fref{fig:predict_m_100}. As we will describe in more detail below, we find the growth depends principally on the effective jet energy efficiency ($\eta_{\rm eff}$). 
After $10^8$ years, a black hole can possibly grow to $10^6 \, {\rm M_\odot}$ when $\eta_{\rm eff}<10^{-5}$ and to $10^7 \, {\rm M_\odot}$ when $\eta_{\rm eff}<10^{-7}$. The lower the feedback mass loading and the lower the jet velocity, the faster the black hole will grow. The transition is sharp; if $\eta_{\rm eff}<10^{-4}$, the black hole grows beyond its initial mass. 

\subsection{The phases of growth}\label{s:phase_of_growth}

To determine the overall amount of mass a BH can accrete, accounting for the jet feedback detailed in this paper, we first need to determine the phases that the system passes through, which we can broadly classify into two cases:

\begin{enumerate}[(i)]
\item {\bf BH growth with $r_{\rm Bondi}<r_{\rm core}$ (constant density)\\}
For jet-based self-regulation in this case, the accretion rate is described by \eeqref{eq:mdotbh} and (\ref{eq:mdotbh_power}), but note that $\dot{M}_{\rm BH}+\dot{M}_{\rm jet}$ should be capped by $\dot{M}_{\rm Bondi}$. If the required $\dot{M}_{\rm jet}$ in  \eeqref{eq:mdotbh} and (\ref{eq:mdotbh_power}) is higher than this value, $\dot{M}_{\rm BH}$ will be at most $\dot{M}_{\rm Bondi}/(1+\eta_{m,fb})$. In both cases, given that the density at the Bondi radius is kept constant in this phase of growth,
    \begin{align}
    \dot{M}_{\rm BH}\propto \dot{M}_{\rm Bondi} \propto M_{\rm BH}^2
    \end{align}

\item {\bf BH growth with $r_{\rm Bondi}>r_{\rm core}$}
When the Bondi radius exceeds the core radius, there are three possibilities:
\begin{enumerate}
\item{\bf Self-regulation without density suppression\\}
In the case of self-regulation, the accretion rate is still described by \Eqref{eq:mdotjet_reg}, but with $\rho_\infty$ replaced by $\rho_{\rm Bondi}(t)$. If density suppression does not occur, $\rho_{\rm Bondi}(r_{\rm Bondi})$ is exactly the initial density at the location. As $\rho_{\rm Bondi}(r_{\rm Bondi})\propto r_{\rm Bondi}^{-2}\propto M_{\rm BH}^{-2}$, we get
\begin{align}
    \dot{M}_{\rm BH}\propto  M_{\rm BH}^2 \rho^\alpha\propto M_{\rm BH}^{-1},
\end{align}
where we follow \Eqref{eq:mdotbh} but adopted $\alpha\sim1.5$, as seen in the simulations.
\item {\bf Failure to regulate\\}
If the required $\dot{M}_{\rm jet}$ in \eeqref{eq:mdotbh} and (\ref{eq:mdotbh_power}) implies that $\dot{M}_{\rm jet}+\dot{M}_{\rm BH}$ is higher than $\dot{M}_{\rm Bondi}$, then, as before, $\dot{M}_{\rm BH}$ will be at most $\dot{M}_{\rm Bondi}/(1+\eta_{m,fb})$. In such a case,
\begin{align}
    \dot{M}_{\rm BH}\propto \dot{M}_{\rm Bondi}\propto  M_{\rm BH}^2 \rho \propto {\rm constant},
\end{align}
so that the accretion rate is independent of $M_{\rm BH}$.
\item {\bf Self-regulation with density suppression\\}
Finally, density suppression occurs when $r_{\rm Bondi}>r_{\rm core}$ and $r_{\rm Bondi}>r_{\rm iso}$. After density suppression, the jet energy flux evolution follows \Eqref{eq:edot_t_after_sup} as $\dot{M}_{\rm BH}=2\dot{E}_{\rm tot, J}/(\eta_{\rm m, fb} V_J^2)$, which barely grows due to the rapid density drop as the cocoon propagates. Note that in this phase, realistically, the propagation of the shock front stops at the terminal radius. After that, another episode of accretion can occur, and we could calculate it by adopting an effective profile with a core radius at the first $r_{\rm terminal}$ and the original density at $r_{\rm terminal}$ as the core density. However, due to the large suppression of the core density, this will not result in significant accretion, and so we neglect such a second or subsequent episode of accretion in \fref{fig:predict_m_100} (and \fref{fig:predict_m_all}, which will be discussed in \sref{s:seed}).
\end{enumerate}    
\end{enumerate}

\begin{figure}
    \centering
    \includegraphics[width=8cm]{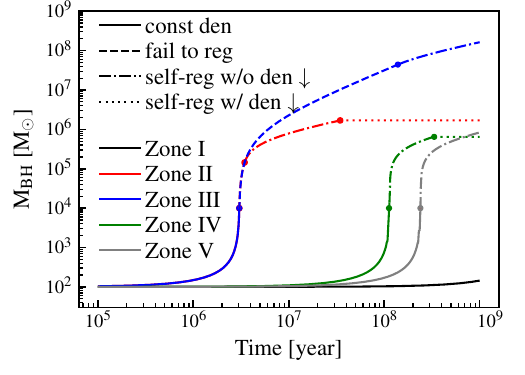}
    \caption{\ksu{Examples of black hole growth with different feedback parameters ($\eta_{\rm m, fb}$, $V_{\rm jet}$) fall into distinct zones: Zone I (0.02,\,$3\times10^4\,{\rm km\, s}^{-1}$), Zone II (0.005,\,$3\times10^3\,{\rm km\, s}^{-1}$), Zone III (0.002,\, $6\times10^2\,{\rm km\,s}^{-1}$), Zone IV (0.1,\,$4\times10^3\,{\rm km\,s}^{-1}$), and Zone V (0.5,\,$3\times10^3\,{\rm km\,s}^{-1}$). The phases of black hole growth that each zone undergoes (see \sref{s:phase_of_growth}) are represented by different line styles, with the phase transitions marked by dots.}}
    \label{fig:example}
\end{figure}

To help guide understanding of these phases and how they connect to the calculated BH growth, we use Roman numerals in \fref{fig:predict_m_100} (and \fref{fig:predict_m_all}) to represent different growth histories, as indicated below. Each Zone indicates the different sequence of growth histories using the notation of the previous enumeration, indicated by arrows\footnote{So for example, Zone II begins with constant density accretion until the BH mass grows such that its Bondi radius reaches the halo core radius, at which point it undergoes an episode of rapid growth due to failed self-regulation until the Bondi radius reaches a sufficiently low density (since $\rho \propto r^{-2}$) that regulation can be reestablished, but without driving density suppression in the core \ksu{(and, if there is sufficient time, the Bondi radius will grow while the isotropization radius shrinks. Once the Bondi radius exceeds the isotropization radius, the BH will ``fee'' density suppression).}}. \ksu{To visualize the sequences described above, \fref{fig:example} presents typical examples of black hole growth with different feedback parameters, each falling into distinct zones. The phases of black hole growth in each zone are represented by different line styles, with phase transitions indicated by dots. }

\begin{enumerate}[Zone I:$\,\,\,\,$]
\item \ksu{The BHs here have high-efficiency feedback, so they never grow beyond $10^4 \, {\rm M_\odot}$.}

\vspace{0.2cm}
{\bf (i) BH growth in constant density}
\vspace{0.4cm}

\item \ksu{The BH here has low-efficiency feedback and goes through the fail-to-regulate phase. The isotropization radius is also small enough that, eventually, density suppression occurs.}

\vspace{0.2cm}
{\bf(i) BH growth in constant density $\\\Rightarrow$ (ii-b) Fail to regulate $\\\Rightarrow$ (ii-c) Self-regulation w/o density suppression $\\\Rightarrow$ (ii-a) Self-regulation with density suppression}
\vspace{0.4cm}

\item \ksu{The BH here has low-efficiency feedback and goes through the fail-to-regulate phase. The isotropization radius is large enough that density suppression has not occurred.}

\vspace{0.2cm}
{\bf (i) BH growth in constant density $\\\Rightarrow$ (ii-b) Fail to regulate $\\\Rightarrow$ (ii-c) Self-regulation w/o density suppression}
\vspace{0.4cm}

\item \ksu{The BH here has intermediate-efficiency feedback and never fails to regulate. The isotropization radius is also small enough that, eventually, density suppression occurs.}

\vspace{0.2cm}
{\bf (i) BH growth in constant density $\\\Rightarrow$ (ii-c) Self-regulation w/o density suppression $\\\Rightarrow$ (ii-a) Self-regulation with density suppression}
\vspace{0.4cm}

\item \ksu{The BH here has intermediate-efficiency feedback and never fails to regulate.The isotropization radius is large enough that density suppression has not occurred.}

\vspace{0.2cm}
{\bf (i) BH growth in constant density $\\\Rightarrow$ (ii-c) Self-regulation w/o density suppression}
\end{enumerate}
Note that for cases with $10^4\,{\rm M_\odot}$ (and $100\,{\rm M_\odot}$ with a cuspy profile), the initial Bondi radius coincides with the core radius, so the black hole has never been through phase (i), and hence there is also no Zone I, as shown in \fref{fig:predict_m_all}. We will discuss this in \sref{s:seed}. \ksu{Black holes with low feedback efficiency (below the light white line in \fref{fig:predict_m_100} and \fref{fig:predict_m_all}) have not reentered the self-regulation phase at the corresponding time.}

\subsection{Dependence on seed BH mass and time}\label{s:seed}
\begin{figure*}
    \centering
    \includegraphics[width=16cm]{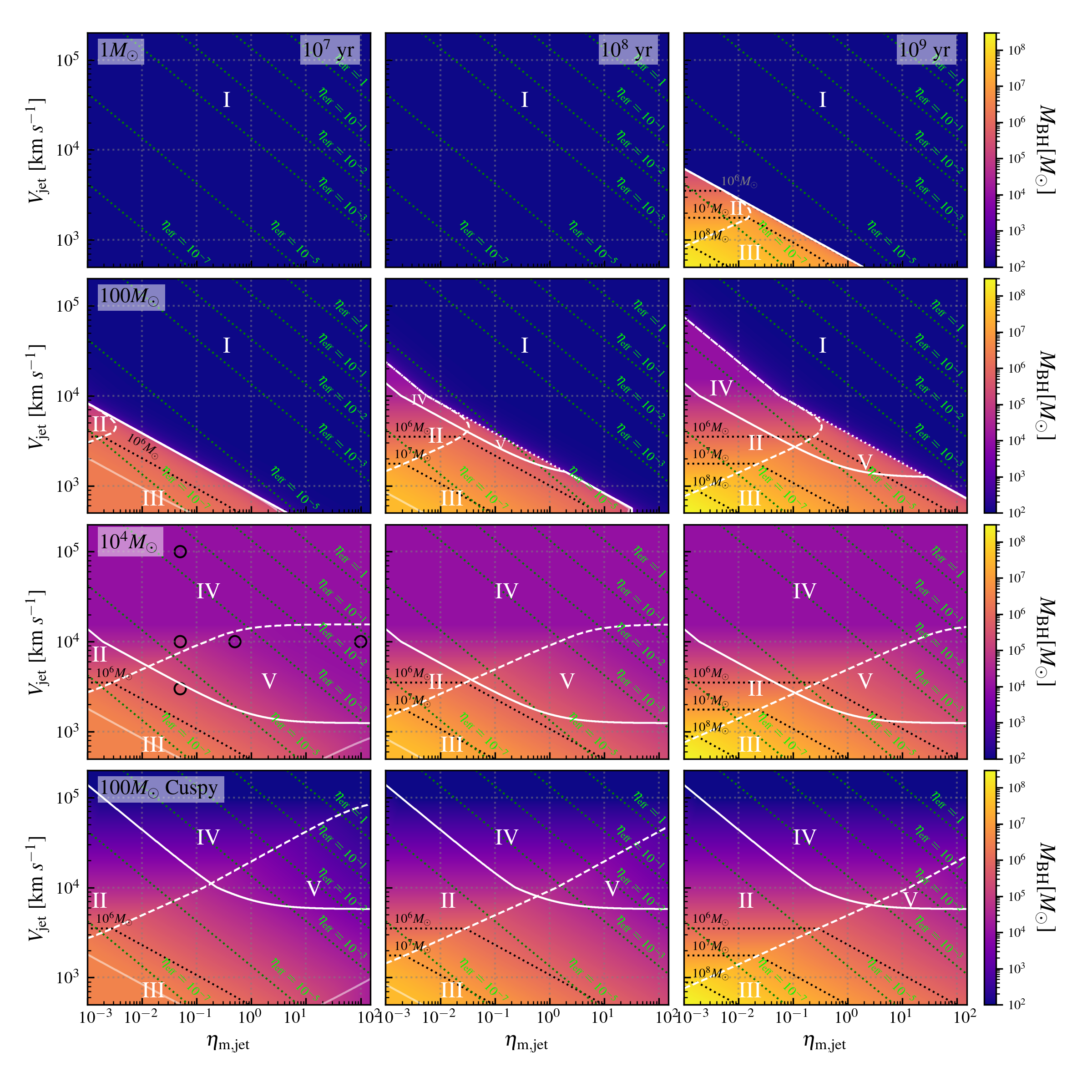}
    \caption{
The prediction of the black hole mass at $10^7$ years (left column), $10^8$ years (center column), and $10^9$ years (right column) after seeding a $1 {\rm M_\odot}$ (1st row), $100 {\rm M_\odot}$ (2nd row), and $10^4 {\rm M_\odot}$ (3rd row) black hole in our fiducial profile of an atomic-cooling halo. The 4th row shows the results for seeding a $100 {\rm M_\odot}$ black hole, assuming a more cuspy profile with $n_{\rm core}=10^9 \, {\rm cm}^{-3}$ and core radius $r_{\rm core}=10^{-2}$ pc. Green dotted lines indicate constant jet energy efficiency, while white lines delineate accretion regions and are described in the text. Seeding a more massive black hole or assuming a cuspy profile of the atomic-cooling halo only affects the final mass for black holes with higher efficiency feedback. }
    \label{fig:predict_m_all}
\end{figure*}

\fref{fig:predict_m_all} shows twelve cases with different seed black hole masses ($1 {\rm M_\odot}$, $100 {\rm M_\odot}$, $10^4 {\rm M_\odot}$), and predictions after $10^7$, $10^8$, and $10^9$ years, and within each panel, the result is shown as a function of jet parameters. The 4th row shows a case with a $100 {\rm M_\odot}$ initial BH, but assuming a more cuspy profile, with $n_{\rm core}=10^9 \, {\rm cm}^{-3}$ and core radius $r_{\rm core}=10^{-2}$ pc. In the cases with $10^4 {\rm M_\odot}$ and $100 {\rm M_\odot}$ in cuspy profiles, the Bondi radius starts at the core radius, so they do not go through growth in the constant density profile (phase (i)).

For the $10^4 {\rm M_\odot}$ case, we have run some simulations to $\sim10^7$ years, allowing an exact comparison with our toy model. The simulation results are marked with circles in the corresponding panel (3rd row, 1st column). The colors in the circles indicate the black hole mass at the end of the simulation, which agrees extremely well with the prediction based on the toy model.

Taking the $100 {\rm M_\odot}$ case as an example, at $10^7$ years, the majority of the parameter space is still in the constant density phase and has not grown much from its initial mass. Only the lowest efficiency cases grow beyond $10^4 {\rm M_\odot}$, mostly due to the ``fail to regulate'' phase. As time evolves, the intermediate efficiency region grows out of the constant density phase and starts to grow, leading to regions IV and V. As time progresses, more of the parameter space grows out of the constant density phase, and regions IV and V become larger.

The parameter space with efficiency lower than the light white line between IV/V and II/III (Zone II, III) undergoes a ``fail to regulate'' phase. As the black hole grows, the Bondi radius increases, and the density at the Bondi radius decreases. However, the required mass flux for regulation scales with $\rho^{1.5}$, while the Bondi accretion rate scales with $n$. As the density decays, self-regulation eventually resumes.

For the highest efficiency parameter space (I), the black hole never grows much beyond its initial mass. For the intermediate efficiency parameter space (IV, V), the black hole never fails to regulate. For the parameter space to the left of the dashed white line, the black hole undergoes a phase of suppressing the density halo due to jet cocoon propagation.

Changing the seed black hole mass mostly affects the high-efficiency part of the parameter space. BHs with high-efficiency feedback grow very slowly, so having a head start with a higher seed mass significantly increases the resulting black hole mass. For the cases with lower efficiency, which do grow to supermassive black holes, changing the seed mass does not change the result, as they reach a black hole mass much larger than $10^4 {\rm M_\odot}$ within a very short time period anyway. The exceptions are the $1 {\rm M_\odot}$ cases within $10^8$ years. Decreasing the seed black hole mass to that level prevents any of the cases from growing beyond $100 {\rm M_\odot}$ within the designated time.

Having a cuspy profile also helps the initial black hole growth (bottom panel). However, black holes with feedback in the low-efficiency parameter space grow to large masses very quickly regardless, so the cuspy profile hardly matters.

We note that, realistically, the final black hole mass for seeding $100 {\rm M_\odot}$ black holes in a cuspy profile should be strictly larger than seeding them in a fiducial profile. We see this is not the case for the upper left corner. This artifact is mostly due to the simplification in the theoretical calculation caused by not considering the core density suppression before the black hole grows to $10^4 {\rm M_\odot}$ when $r_{\rm Bondi}$ reaches $r_{\rm core}=1$ pc. Realistically, such suppression can happen earlier when the cocoon isotropizes, and the shock front already propagates beyond $r_{\rm core}$. Therefore, we slightly overestimate the final black hole mass for the upper left corner of region IV for the $100 {\rm M_\odot}$ fiducial profile case.

\subsection{Maximum feedback efficiency for forming supermassive black hole}
\label{sec:maximum_feedback_efficiency}
\begin{figure}
    \centering
    \includegraphics[width=8cm]{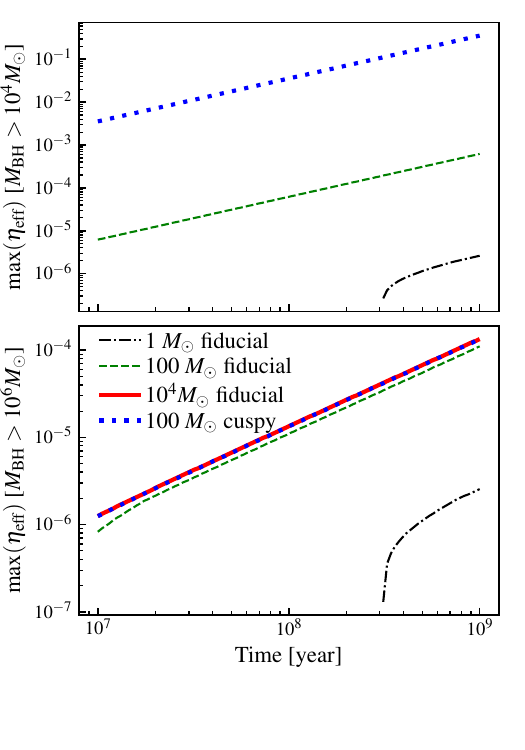}
    \caption{The maximum efficiency for seeding different halo masses in different gas profiles, allowing the black hole to grow to intermediate mass black holes ($10^4-10^6 {\rm M_\odot}$, upper) or supermassive black holes ($>10^6 {\rm M_\odot}$, lower) assuming $V_{\rm Jet}>500 km s^{-1}$, is shown as a function of time. For most cases, except for a 1 ${\rm M_\odot}$ seed, an efficiency smaller than $\eta_{\rm eff}=(10^{-6}, 10^{-5}, 10^{-4})$ is required for the black hole to grow to supermassive status by $t=(10^7, 10^8, 10^9)$ years. Having cuspier density profiles and increasing the seed black hole mass can increase the feedback efficiency upper bound for forming intermediate mass black holes by orders of magnitude. }
    \label{fig:predict_eff}
\end{figure}

\fref{fig:predict_eff} shows the maximum jet efficiency ($\eta_{\rm eff}$ defined in \Eqref{eq:e_eff}) that will allow a black hole to grow to intermediate mass scale ($10^4-10^6 {\rm M_\odot}$; top panel) or supermassive black holes ($>10^6 {\rm M_\odot}$; top panel) as a function of time, for different seed masses and halo properties (assuming $V_{\rm Jet}>500\, {\rm km\, s}^{-1}$). For most cases, except for the 1 ${\rm M_\odot}$ seed, an efficiency smaller than $\eta_{\rm eff}=(10^{-6}, 10^{-5}, 10^{-4})$ is required for the black hole to grow to supermassive size within $t=(10^7, 10^8, 10^9)$ years.

Seeding a 1 ${\rm M_\odot}$ black hole requires more than 2 orders of magnitude lower efficiency to grow to supermassive size within the same time frame, and it is only possible after $3\times 10^8$ years. Having a cuspy profile or seeding a $10^4 {\rm M_\odot}$ black hole slightly increases the efficiency upper bound for supermassive black holes, by a similar extent, within a factor of 2.

However, having cuspier density profiles and increasing the seed black hole mass can increase the allowed feedback efficiency for forming intermediate mass black holes by orders of magnitude. Both of these factors give the black hole growth a head start, allowing black holes with feedback over a larger parameter space to grow to $>10^4 {\rm M_\odot}$.

\section{Discussion}\label{sec:discussion}

\subsection{Limitations of the model}\label{sec:limitation}
In this work, we focus on setups with constant initial temperature and an idealized density profile: a core with constant density within 1 pc, transitioning to an $n\propto r^{-2}$ profile beyond that. We neglect any initial gas motion, such as turbulence or rotation. While turbulence arises after the jet is launched, we do not include any other causes of gas motion. Realistically, both rotation and turbulence can significantly hinder accretion. We also neglect the presence of magnetic fields, which can further suppress accretion.

Our simulations are run for a finite amount of time, capturing only a glimpse of the various phases of black hole growth. Although our predictions for black hole mass after a time longer than the simulation runtime are based on physical understanding and a toy model that faithfully describes the simulation results, there is a significant extrapolation that needs verification in future work.

We focus solely on jets, which are highly collimated mechanical feedback mechanisms. Realistically, there can be other forms of AGN feedback, such as winds with broader opening angles and radiative feedback, which can result in very different cocoon propagation.

Most importantly, we neglect other galactic processes like star formation and stellar feedback, which can further suppress black hole accretion. We also do not consider the later growth of the atomic-cooling halo or mergers. As a result, our predicted black hole mass should be viewed as an upper bound for black holes accreting within a single atomic halo. We emphasize that there are many other channels through which black holes can grow, and we are not able to constrain those in this work.

\ksu{We also anchor the black hole at the center of the atomic-cooling halo. In reality, the black hole seed may `'wander'' away from the halo center and may not be located in the densest region \citep{2024arXiv240517975R}.}

\subsection{Observational implication}
Despite the limitations discussed in  \sref{sec:limitation}, this work provides a strong constraint on the channel of black hole growth that relies on accreting gas within a single atomic-cooling halo. Recent observations with JWST have identified a population of supermassive black holes (SMBHs) at relatively high redshifts, beyond $z \sim 6$ and even beyond $z \sim 10$ \citep[e.g.,][]{2024Natur.627...59M,2023ApJ...953L..29L,2023Natur.619..716C,2023ApJ...959...39H,2023ApJ...954L...4K,2023ApJ...942L..17O,2023A&A...677A.145U,2024MNRAS.531.4584S}. Assuming an atomic-cooling halo forms at $z \sim 12$ and a black hole is seeded immediately, $10^8$ and $10^9$ years later would correspond to $z \sim 10$ and $z \sim 5$, respectively.

As shown in \fref{fig:predict_m_all} and \fref{fig:predict_eff}, even a small amount of collimated mechanical feedback with relatively low efficiency can significantly hinder black hole growth. If we rely on this channel to explain the SMBHs observed by JWST, the feedback efficiency should be $\ll 10^{-4}$, and even lower to form those at $z > 10$. Heavy seeds do not help in this scenario, as most of the time spent reaching supermassive black hole status occurs when the black hole mass exceeds $10^4 {\rm M_\odot}$. Alternatively, the black holes observed at high redshift could form in even denser environments than a typical atomic-cooling halo or grow via other channels.

\section{Conclusion}\label{sec:summary}
In this work, we have utilized a set of idealized simulations with black holes of different masses seeded in various gas environments mimicking the centers of atomic-cooling halos. Based on these simulations, we provide a toy model describing the propagation of jet cocoons and their resulting regulation of black hole growth. Using this toy model, we predict black hole mass as a function of time, assuming different seed masses. We found that even with relatively low feedback efficiency, the central density profile of the atomic-cooling halo can be largely suppressed after the first episode of jet cocoon propagation. After this density suppression, black hole growth essentially stops. Thus, very low mechanical feedback efficiency is required to form a supermassive black hole (SMBH) at high redshift if relying on feeding a black hole by a single atomic-cooling halo. We summarize our conclusions as follows:
\begin{itemize}
\item We confirm the toy model presented in \cite{2023MNRAS.520.4258S} for jet cocoon propagation across various black hole masses. The propagation of the jet cocoon in the jet direction is governed by momentum conservation, while the lateral expansion is governed by energy conservation due to the pressure in the cocoon. Eventually, the lateral velocity increases and becomes comparable to the velocity in the jet direction at the isotropization radius ($r_{\rm iso}$). Beyond this radius, the jet cocoon becomes an energy-driven isotropic bubble. Lower jet velocity and higher density result in a larger isotropization radius.
\item We confirm that despite different comparisons of the isotropization radius ($r_{\rm iso}$) and the Bondi radius ($r_{\rm Bondi}$), it is always the isotropic component of the cocoon momentum that is regulated by the inflowing momentum flux, assuming a Bondi solution at the Bondi radius.
\item Super-Eddington accretion can occur when the black hole mass reaches approximately $\sim100 {\rm M_\odot}$ in cases with high density, low temperature, and/or low feedback efficiency.
\item As the jet cocoon isotropizes and propagates beyond the core radius ($r_{\rm core}$), it can significantly suppress the density profile at the center of the atomic-cooling halo. This density suppression occurs from the isotropization radius ($r_{\rm iso}$) to the current location of the cocoon shock front, making the density roughly constant within that range. If the Bondi radius ($r_{\rm Bondi}$) falls within this radial range, the black hole accretion is affected by the density suppression.
\item The cocoon will propagate to the terminal radius, where the integrated cooling of the gas within the jet cocoon balances the integrated jet energy flux, stalling its growth.
\item Despite the density suppression, we find that the isotropic component of the outflowing momentum flux is still regulated to match the Bondi inflowing momentum at the Bondi radius, accounting for the real-time density at that radius. This results in a secularly evolving regulation scenario.
\item Based on an analytic model inspired and calibrated by the simulations, we provide a prediction of black hole mass growth as a function of time and black hole seed mass, assuming accretion from a single atomic-cooling halo. To form a supermassive black hole within $10^8$ and $10^9$ years, we require a jet efficiency of $\eta < 10^{-5}$ and $\eta < 10^{-4}$, respectively.
\item For black holes with feedback in the parameter space that allows growth to a supermassive black hole, most of the time is spent when the black hole mass exceeds $10^4 {\rm M_\odot}$. Therefore, having a heavier seed or assuming a cuspier profile does not significantly increase the efficiency upper bound for supermassive black hole formation.
\item On the other hand, having a heavier seed or a cuspier profile provides a head start for black hole growth, assuming high feedback efficiency. This allows black holes with a larger feedback parameter space to reach intermediate-mass black holes.
\item We identified several phases of black hole growth (\fref{fig:predict_m_100} and \fref{fig:predict_m_all}). For the feedback parameter space that allows the formation of supermassive black holes, most of the accreted mass occurs during the growth phase, where the very low-efficiency feedback fails to regulate the Bondi accretion.

\end{itemize}

We reemphasize that the above conclusions only apply to black hole growth relying on feeding within a single typical atomic halo, providing constraints specific to this growth channel. We also neglect stellar physics and magnetic fields, which may further hinder black hole growth. Our predictions should be viewed as an upper bound for black hole accretion. We leave these other aspects for future study.

\vspace{-0.2cm}
\acknowledgments
KS acknowledges support from the Gordon and Betty Moore Foundation and the John Templeton Foundation via grants to the Black Hole Initiative at Harvard University.   GLB acknowledges support from the NSF (OAC-1835509, AST-2108470), a NASA TCAN award, and the Simons Foundation.
RSS was supported by the Simons Foundation through the Flatiron Institute. ZH acknowledges support from NSF grant AST-2006176.
Numerical calculations were run on the Flatiron Institute cluster ``popeye'' and ``rusty'' Numerical calculations were run on the Flatiron Institute cluster ``popeye'' and ``rusty'', Frontera with allocation AST22010, and Bridges-2 with Access allocations TG-PHY220027 \& TG-PHY220047. This work was performed in part at Aspen Center for Physics, which is supported by National Science Foundation grant PHY-2210452. We also thank the FIRE and the LtU collaborations.
\vspace{0.3cm}
\section*{Data Availability }
The data supporting the plots within this article are available on reasonable request to the corresponding author. A public version of the GIZMO code is available at \href{http://www.tapir.caltech.edu/~phopkins/Site/GIZMO.html}{\textit{http://www.tapir.caltech.edu/$\sim$phopkins/Site/GIZMO.html}}.

\bibliographystyle{mnras}
\bibliography{mybibs}

\label{lastpage}

\end{document}